\documentclass[aps,prd,reprint,a4paper,groupedaddress,nofootinbib,longbibliography,floatfix]{revtex4-2}
\usepackage{iftex}
\iftutex
\usepackage[UTF8,heading=false,scheme=plain,fontset=fandol]{ctex}
\fi
\usepackage[T1]{fontenc}
\usepackage{amssymb,bm,mathtools}
\usepackage{booktabs,tabularx,array,graphicx}
\usepackage[caption=false]{subfig}
\usepackage{needspace}
\usepackage{placeins}
\usepackage[colorlinks=true,linkcolor=blue,citecolor=blue,urlcolor=blue,filecolor=blue]{hyperref}
\graphicspath{{figures/}}

\hypersetup{hypertexnames=false}
\ifxetex
\AddToHook{shipout/firstpage}{}
\fi

\newcommand{\iu}{\mathrm{i}}
\newcommand{\dd}{\mathrm{d}}

\newcommand{\Rcal}{\mathcal R}
\newcommand{\Kcal}{\mathcal K}
\newcommand{\Ain}{A_{\mathrm{in}}}
\newcommand{\Aout}{A_{\mathrm{out}}}

\newcolumntype{Y}{>{\raggedright\arraybackslash}X}
\newcommand{\GridCell}[1]{\begin{tabular}[c]{@{}c@{}}#1\end{tabular}}
\newcommand{\GridTableSetup}{%
 \small
 \setlength{\tabcolsep}{2em}
 \setlength{\arrayrulewidth}{0.45pt}
 \renewcommand{\arraystretch}{1.30}
}
\begin{document}
\title{\boldmath Rotational Splitting of Reflectionless Matching and Virtual Absorption in a Semiopen BTZ Cavity}
\newcommand{\ManuscriptAffiliations}{\parbox{6.7in}{\centering
$^{a}$School of Fundamental Physics and Mathematical Sciences, Hangzhou Institute for Advanced Study, University of Chinese Academy of Sciences,\par
Hangzhou 310024, People's Republic of China\par\vspace{2pt}
$^{b}$School of Physics and Telecommunications Engineering, Zhoukou Normal University, Zhoukou 466001, China\par\vspace{2pt}
$^{c}$Department of Physics, Key Laboratory of Low Dimensional Quantum Structures and Quantum Control of Ministry of Education, Synergetic Innovation Center for Quantum Effects and Applications, Hunan Normal University, Changsha, 410081, Hunan, China\par\vspace{2pt}
$^{d}$Institute of Interdisciplinary Studies, Hunan Normal University, Changsha, Hunan 410081, China\par\vspace{2pt}
$^{e}$Department of Physics, Huaihua University, Huaihua, Hunan 418008, China}}
\author{Sheng Long$^{a}$}
\author{Manman Sun$^{b}$}
\author{Qin Tan$^{c,d,*}$}
\author{Yu Shi$^{e,*}$}
\affiliation{\ManuscriptAffiliations}
\preprint{RBTZ-TTM-001}

\begin{abstract}
We study complex frequency reflection zeros of a scalar field in a finite rotating BTZ region. The region is bounded by a partially reflecting near horizon wall and connected to an exterior propagation channel. We use exact radial solutions to determine the conditions for zero reflection and test their finite time behavior through wave packet evolution. Rotation splits the reflection zeros in the $m=\pm1$ sectors. For a representative pair of branches, there is a range of reflection coefficients for a passive wall in which the co-rotating incident wave must grow in time, while the counter-rotating one must decay. In the near horizon limit, the magnitude of the wall reflection coefficient mainly controls the growth or decay rate, while its phase mainly controls the oscillation frequency. Exterior matching introduces corrections to these relations. We examine four truncated wave packets, including one tuned case and three reference cases. Tuning reduces the reflected energy fraction from $(3.5$--$7.6)\times10^{-4}$ to $1.8\times10^{-4}$, with little change in wall absorption or the peak energy stored in the cavity. We also find that two co-rotating branches merge at a second-order exceptional point (EP2). A near horizon expansion determines its location and the small wall reflectivity required for its formation. Under perturbations, the reflection zeros exhibit square-root splitting, while the poles of the local-source Green's function do not merge and its response remains linear. These results distinguish scattering zeros from the intrinsic resonances of the system.
\end{abstract}

\maketitle
\begingroup
\renewcommand{\thefootnote}{\fnsymbol{footnote}}
\footnotetext[1]{Corresponding authors: Qin Tan
  (\href{mailto:tanqin@hunnu.edu.cn}{tanqin@hunnu.edu.cn}),\newline
  Yu Shi (\href{mailto:shiyu@hhtc.edu.cn}{shiyu@hhtc.edu.cn}).}
\endgroup
%

\section{Introduction}
\label{sec:introduction}


Quasinormal modes (QNMs) are the poles of the black hole Green's function: the real and imaginary parts of their complex frequencies give, respectively, the oscillation frequency and the damping rate of the ringdown~\cite{Kokkotas:1999bd,Berti:2009kk, Konoplya:2011qq}, and hence form the basis for spectroscopic tests of merger remnants~\cite{Dreyer:2004spectroscopy,Berti:2006spectroscopy}; meanwhile, the residues of the QNMs determine the strength with which a given source excites each mode~\cite{Berti:2006excitation}. In asymptotically anti–de Sitter (AdS) spacetimes, the same set of QNM poles also governs the retarded boundary correlators~\cite{Horowitz:1999jd,Kovtun:2005ev,Son:2002sd}. Unlike the eigenvalue problem above, reflection zeros answer a different kind of scattering question: which incident waves undergo no reflection at all during scattering? If a partially reflecting surface is introduced near the location where a horizon would otherwise form~\cite{Cardoso:2016horizon,Burgess:2018echo}, this question becomes connected to a class of delayed-signal models; in such models, propagation, external scattering, and surface reflection together produce the delayed signal~\cite{Mark:2017echo,Maggio:2019echo}.


For a specified incident direction, the condition of “vanishing reflection” defines the totally transmissive modes in the study of Schwarzschild and Kerr perturbations~\cite{MaassenVanDenBrink:2000,Cook:2019ttm,Cook:2023ttm}. Reflectionless scattering theory formulates this condition in terms of complex frequency reflection zeros~\cite{Sweeney:2020rsm}; in a dissipative single channel system, zeros lying on the real frequency axis correspond to coherent perfect absorption~\cite{Chong:2010cpa}. When the zeros move off the real axis, their positions further dictate the growth or decay of the incident wave in time: an exponentially growing incident wave can exhibit coherent virtual absorption~\cite{Baranov:2017,Radi:2020vcc}. Recent studies have, on the one hand, linked reflection zeros to echo phenomena~\cite{Rosato:2025greybody,Qian:2025rsm}, and on the other hand examined black hole scattering as well as tailored incidence in the semi-open Schwarzschild--de Sitter geometry~\cite{Tuncer:2025dnp,Wu:2026vam}.


Rotation breaks the equivalence between the two sectors of opposite azimuthal angular momentum; the same effect appears in the echoes of rotating bodies and in the totally transmissive modes of acoustic black holes~\cite{Maggio:2019echo,Yu:2026dbm}. The rotating BTZ geometry admits exact solutions of the radial equation for a scalar field, so that its absorption problem can be treated analytically~\cite{Banados:1992gq,Cardoso:2001hn,Birmingham:2001pj,Birmingham:1997absorption}. In this setting, the radius-dependent combination $\omega+mN^\varphi(r)$ allows us to distinguish the distinct roles of the wall, the propagation, and the external matching: the rotational correction cannot simply be reduced to a constant frequency shift. For this purpose, we retain a finite BTZ interval and connect it to an external channel with definite incident and reflected amplitudes, without imposing a boundary condition at AdS infinity~\cite{Ishibashi:2004wx,Dappiaggi:2017pbe}.

%

The second question arises when two reflection zeros coalesce at an exceptional point (EP): the two then share a single eigenfunction, and their frequency splitting takes a square-root form~\cite{Kato:1995,Heiss:2012qla}. Previous studies have examined EPs in black hole QNM spectra~\cite{Motohashi:2024fwt,PanossoMacedo:2025xnf,Cavalcante:2024ep,Nakamoto:2026nariai}, including the response of the rotating BTZ~\cite{Long:2026btzep} and EPs in absorption zeros~\cite{Sweeney:2019chiral}. Together with studies on the sensitivity of totally transmissive modes~\cite{Zhou:2025xdo,Dong:2026bumps}, these results prompt us to investigate where the smooth tuning of reflection zeros breaks down.

After defining the scattering problem in Sec.~\ref{sec:ports}, we obtain three sets of results. First, rotation separates the critical wall coefficients corresponding to the paired $m=\pm1$ reflection zeros; between the two thresholds of a representative paired branch, one matched incident wave grows with time while the other decays (Sec.~\ref{sec:critical-coupling}). Second, for a truncated wave packet, tuning to the reflection zero reduces the reflected fraction from $(3.5$--$7.6)\times10^{-4}$ to $1.8\times10^{-4}$, while the wall absorption and the peak energy in the region remain almost unchanged (Sec.~\ref{sec:time-domain}). Third, the two corotating reflection-zero branches merge at an EP, whose location and the requisite small wall reflection coefficient can both be obtained from a near-horizon inverse mapping; we also compare their square-root splitting with the response induced by a localized source (Sec.~\ref{sec:reflection-zero-ep}).

\section{Scattering in a rotating BTZ cavity}
\label{sec:ports}

\begin{figure*}[tp]
	\begin{minipage}{6.5in}
		\centering
		\includegraphics[width=\textwidth]{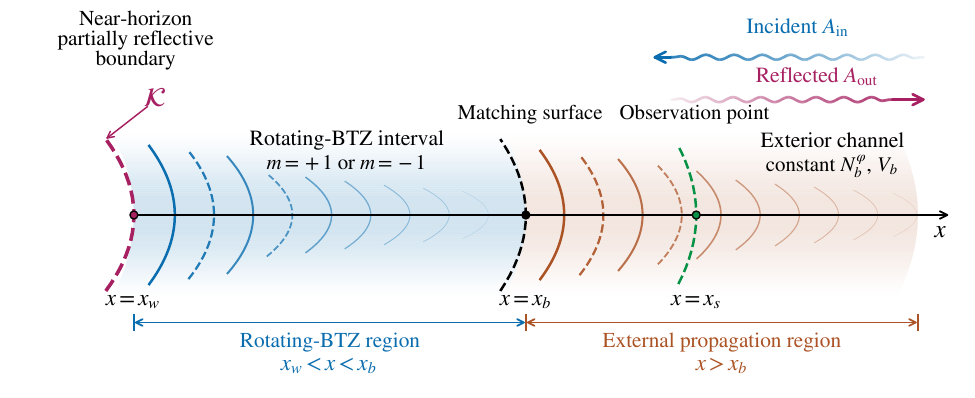}
		\caption{The BTZ interval extends from the partially
			reflecting wall at $x_w$ to the exterior matching surface at
			$x_b$, with exterior amplitudes $\Ain$ and $\Aout$.
			The region used for energy integration, $[x_w,x_s]$ with $x_s=5$,
			includes part of the exterior channel.}
		\label{fig:semi-open-cavity}
	\end{minipage}
\end{figure*}

We study scattering through a finite interval of the rotating BTZ geometry (the BTZ interval), bounded by a partially reflecting inner wall and by an exterior matching surface that connects to an idealized propagation channel (Fig.~\ref{fig:semi-open-cavity}). At a reflection zero (a frequency at which the reflection vanishes), a purely incident exterior wave, after propagating through the cavity, satisfies the wall condition.

The rotating BTZ metric, with AdS radius $L$ and horizon radii
$r_+>r_-\geq0$, is
\begin{equation}
	\begin{aligned}
		\dd s^2&=-N^2\dd t^2+N^{-2}\dd r^2
		+r^2(\dd\varphi+N^\varphi\dd t)^2,\\
		N^2&=\frac{(r^2-r_+^2)(r^2-r_-^2)}{L^2r^2},
		\qquad
		N^\varphi=-\frac{r_+r_-}{Lr^2}.
	\end{aligned}
	\label{eq:ttm-btz-metric}
\end{equation}
For a scalar field satisfying $(\Box-\mu^2)\Phi=0$, the ansatz
$\Phi=\Psi(r)e^{-\iu\omega t+\iu m\varphi}/\sqrt r$, with
$\mu^2L^2=\nu^2-1$ and $0<\nu<1$, gives
\begin{equation}
	\Psi_{xx}+\left[(\omega+mN^\varphi)^2-V_m(r)\right]\Psi=0.
	\label{eq:ttm-schrodinger}
\end{equation}
Here $x$ is the tortoise coordinate and $\dd x/\dd r=N^{-2}$;
$x\to-\infty$ as the outer horizon is approached, and $x\to0$ at
the AdS boundary of the untruncated geometry.
We introduce the dimensionless variables $w=\omega L^2/r_+$,
$a=r_-/r_+$, $q=mL/r_+$ and $z=(r^2-r_+^2)/(r^2-r_-^2)$;
the horizon angular velocity is $\Omega_H=r_-/(Lr_+)$.

The BTZ interval is restricted to $z_w\leq z\leq z_b<1$.
At the near-horizon wall, we impose
\begin{equation}
	\begin{aligned}
		\mathcal B_w^{(Z)}[\Psi]
		&=\left[\Psi_x+\iu Z_ws_w\Psi\right]_{x_w}=0,\\
		Z_w&=\frac{1-\Kcal}{1+\Kcal},\qquad
		s_w=\omega+mN_w^\varphi.
	\end{aligned}
	\label{eq:ttm-wall-impedance}
\end{equation}
In the local wave decomposition associated with this Robin condition,
$\Kcal$ is the wall amplitude reflection coefficient and $|\Kcal|^2$ is
the flux reflectivity. We use a real, frequency-independent $\Kcal$
for wave packets and allow complex $\Kcal$ in reflection-phase
studies. A frequency-dependent membrane response provides a possible
extension \cite{Chakraborty:2022membrane}.

For $x>x_b$, freezing $N^\varphi$ and $V_m$ at their matching-surface
values defines the exterior channel,
\begin{equation}
	\begin{aligned}
		\Psi&=\Ain e^{-\iu k_b(x-x_b)}
		+\Aout e^{\iu k_b(x-x_b)},\\
		k_b^2&=(\omega+mN_b^\varphi)^2-V_b.
	\end{aligned}
	\label{eq:ttm-lead-decomposition}
\end{equation}
We take $k_b>0$ on the real-frequency propagation branch and
analytically continue this branch to complex frequencies.
Reflectionless incidence, $\Aout=0$ with $\Ain\ne0$, requires
\begin{equation}
	\mathcal B_b[\Psi]=[\Psi_x+\iu k_b\Psi]_{x_b}=0.
	\label{eq:ttm-condition}
\end{equation}

Let $j$ and $o$ denote the solutions obtained by continuing
unit-amplitude, purely incident and purely outgoing exterior waves
inward, respectively.
Their scaled wall residuals give
\begin{equation}
	\begin{aligned}
		\Rcal&=\frac{\Aout}{\Ain}=-\frac{D}{E},\\
		D&=(1+\Kcal)\mathcal B_w^{(Z)}[j],\qquad
		E=(1+\Kcal)\mathcal B_w^{(Z)}[o].
	\end{aligned}
	\label{eq:multimode-real-R}
\end{equation}
Along regular analytic branches away from $k_b=0$,
$D=0$ with $E\ne0$ corresponds to reflectionless incidence, whereas
$E=0$ with $D\ne0$ corresponds to purely outgoing resonances
($\Ain=0$). These conditions are analogous, respectively, to the
TTM and QNM boundary prescriptions in black-hole scattering
\cite{Andersson:1994ttm,MaassenVanDenBrink:2000,Cook:2014cta,Cook:2019ttm}.
Together, the frequency-dependent equation and the boundary
conditions define a non-self-adjoint nonlinear spectral problem.
The potential, exact radial solutions and scattering amplitudes
are given in Supplement~A.
This scattering problem differs from the BTZ brick-wall models,
which impose an inner Dirichlet wall and normalizability at
AdS infinity \cite{Das:2024brickwall,Belokon:2026brickwall}.

A complex reflection zero fixes both the oscillation frequency and
the incident envelope through
$e^{-\iu\omega t}
=e^{-\iu\operatorname{Re}(\omega)t}
e^{\operatorname{Im}(\omega)t}$.
Upper-half-plane zeros select exponentially growing incident waves,
termed virtual absorption modes.
For real $\omega$ and $\Kcal$, with $|\Kcal|<1$, $k_b>0$ and
$s_w>0$, we have $Z_w>0$, and flux conservation then gives
\begin{equation}
	k_b(|\Ain|^2-|\Aout|^2)
	=Z_ws_w|\Psi_w|^2.
	\label{eq:critical-flux-balance}
\end{equation}
A real-frequency reflection zero therefore yields perfect
monochromatic absorption at the wall.
Finite-duration wave packets are studied in
Sec.~\ref{sec:time-domain}.
Reference parameters are listed in Table~\ref{tab:ttm-parameters}.

\begin{table}[t]
	\centering
	\caption{Reference parameters. Changes are stated where used.}
	\label{tab:ttm-parameters}
	\begingroup\GridTableSetup
	\begin{tabular}{|c|c|}
		\hline
		Physical quantity & Reference value \\
		\hline
		Geometry & \GridCell{$L=1,\ r_+=2$\\$a=0.6$} \\
		\hline
		Scalar field & $\nu=0.65$ \\
		\hline
		Azimuthal sectors & $m=\pm1$ \\
		\hline
		Inner boundary & $z_w=10^{-12}$ \\
		\hline
		Matching surface & $z_b=0.35$ \\
		\hline
		\GridCell{Wall reflection\\coefficient $\Kcal$}
		& \GridCell{$-0.3$ (spectral catalog)\\$-0.015$ (wave packets)} \\
		\hline
		Observation point & $x_s=5$ \\
		\hline
	\end{tabular}
	\endgroup
\end{table}

Unless stated otherwise, spin scans keep $z_w$ and $z_b$ fixed,
so both the circumferential matching radius and the cavity length in
tortoise coordinates vary with spin. Frequency shifts therefore
include changes in both radial propagation and coupling to the
exterior channel. Numerical methods and coverage checks are
summarized in Supplement~C.1.

\section{Rotation and wall control of reflection zeros}
\label{sec:critical-coupling}
\label{sec:spectral-flow}

\begin{figure*}[tp]
	\begin{minipage}{6.5in}
		\centering
		\subfloat[Critical wall reflection magnitude.
		\label{fig:critical-coupling-regime:a}]{%
			\includegraphics[width=0.485\textwidth]{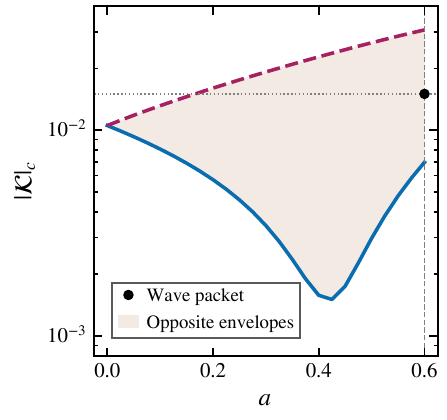}}\hfill
		\subfloat[Growth and decay rates at $\Kcal=-0.015$.
		\label{fig:critical-coupling-regime:b}]{%
			\includegraphics[width=0.485\textwidth]{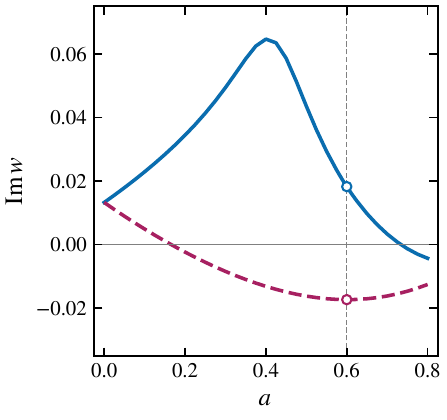}}
		\par\smallskip
		\includegraphics{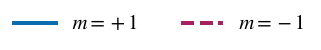}
		\caption{Critical coupling and imaginary frequencies for $n=5$.
			Shading marks the interval between $|\Kcal|_{c,+}$ and
			$|\Kcal|_{c,-}$. The black dot and open circles identify
			the wave-packet parameters at $a=0.6$; the horizontal
			line in (b) marks $\operatorname{Im}w=0$.}
		\label{fig:critical-coupling-regime}
	\end{minipage}
\end{figure*}

Rotation splits the reflection zeros for $m=\pm1$
through the radial term $2\omega mN^\varphi(r)$, which changes
sign with $m$ and produces different propagation amplitudes
and phases. At $a=0$, the two sectors are degenerate because
the radial equation and boundary conditions depend only on
$m^2$. Continuation from this common limit identifies the
paired branches. We label them by $n$ in order of increasing
nonrotating frequency and use $n=5$ as a representative pair;
the correspondence to the numerical catalog is given in the
Supplemental Material. The radial dependence of $N^\varphi$
means that rotation changes propagation throughout the interval,
not just the frequency measured at the wall.

A real-frequency reflection zero corresponds to
constant-envelope incidence and, for a passive wall, critical
coupling~\cite{Chong:2010cpa,Sweeney:2020rsm}.
Figure~\ref{fig:critical-coupling-regime} shows distinct
critical wall reflection magnitudes:
$|\Kcal|_{c,+}\simeq0.007$ and
$|\Kcal|_{c,-}\simeq0.031$ at $a=0.6$.
The symmetry $(a,m)\mapsto(-a,-m)$ implies
$(w_c,\eta_c)_+(a)=(w_c,\eta_c)_-(-a)$, where
$\eta_c=\log|\Kcal|_c$, for regular, locally unique
critical crossings. Consequently,
\begin{equation}
	\begin{aligned}
		\Delta w_c
		&=w_{c,+}-w_{c,-}=2aw_1+O(a^3),\\
		\Delta \eta_c
		&=\log\frac{|\Kcal|_{c,+}}{|\Kcal|_{c,-}}
		=2a\eta_1+O(a^3).
	\end{aligned}
	\label{eq:small-spin-splitting}
\end{equation}
For $n=5$, $w_1\simeq0.49$ and $\eta_1\simeq-2.39$,
as determined by radial matching
(Supplement~A.2).
The resulting interval
\begin{equation}
	|\Kcal|_{c,+}<|\Kcal|<|\Kcal|_{c,-}
	\label{eq:selective-window}
\end{equation}
selects an exponentially growing corotating wave and a decaying
counterrotating wave under the $e^{-\iu\omega t}$
convention. The reference choice $\Kcal=-0.015$ lies
in this interval and is used for the wave-packet comparisons.

\begin{figure*}[tp]
	\begin{minipage}{6.5in}
		\centering
		\subfloat[$m=+1$, $\Kcal=-0.3$.
		\label{fig:multimode-spin:a}]{%
			\includegraphics[width=.48\textwidth]{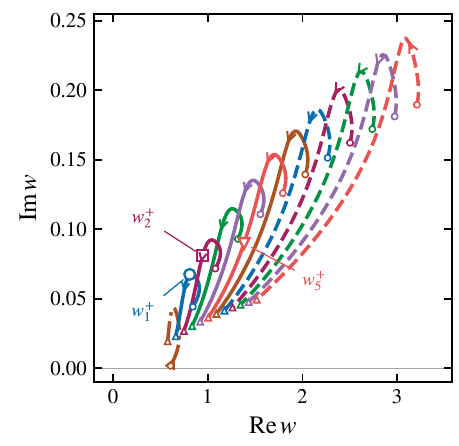}}\hfill
		\subfloat[$m=-1$, $\Kcal=-0.3$.
		\label{fig:multimode-spin:b}]{%
			\includegraphics[width=.48\textwidth]{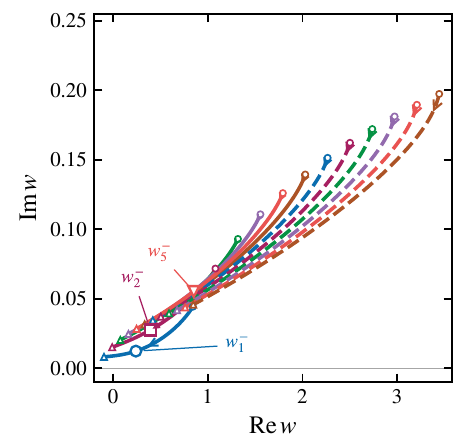}}
		\par\medskip
		\subfloat[$m=+1$, $\Kcal=-0.015$.
		\label{fig:multimode-spin:c}]{%
			\includegraphics[width=.48\textwidth]{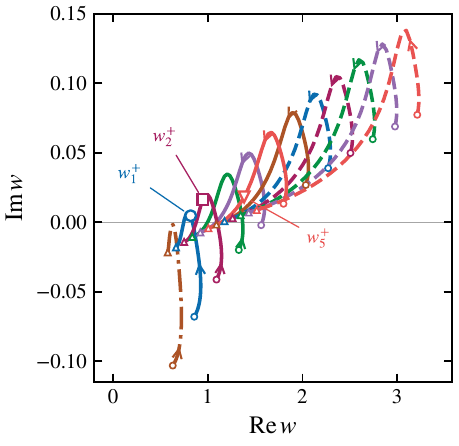}}\hfill
		\subfloat[$m=-1$, $\Kcal=-0.015$.
		\label{fig:multimode-spin:d}]{%
			\includegraphics[width=.48\textwidth]{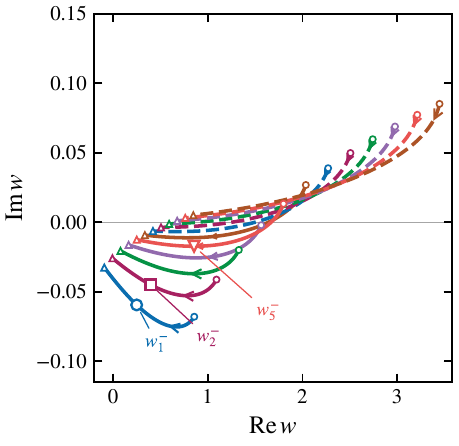}}
		\par\smallskip
		\includegraphics{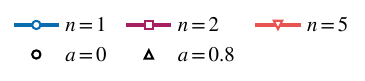}
		\caption{Spin evolution of twelve tracked reflection zeros
			per panel. Arrows indicate increasing $a$; small circles
			and triangles mark the endpoints $a=0,0.8$, and larger
			markers identify reference pairs at $a=0.6$.
			The corotating $n=1,2$ branches form the EP pair,
			while $n=5$ is used for the wave packets.
			Labels are fixed along each trajectory.}
		\label{fig:multimode-spin}
	\end{minipage}
\end{figure*}

The imaginary parts of the zeros in Fig.~\ref{fig:multimode-spin}
vary differently with spin on different radial branches.
At $\Kcal=-0.015$, four lower corotating branches cross the
real axis twice: their reflectionless incident waves change
from decaying to growing and then back to decaying as $a$
increases. The $n=5$ branch crosses once, while several higher
branches do not cross. The real parts also evolve differently
between sectors: corotating frequencies can turn, whereas the
counterrotating frequencies decrease over the displayed range.
The incident frequency and growth rate must therefore be chosen
for a particular branch, even when the same wall is used.

\begin{figure*}[tp]
	\begin{minipage}{6.5in}
		\centering
		\subfloat[Critical wall reflection magnitude.
		\label{fig:multimode-critical:a}]{%
			\includegraphics[width=0.485\textwidth]{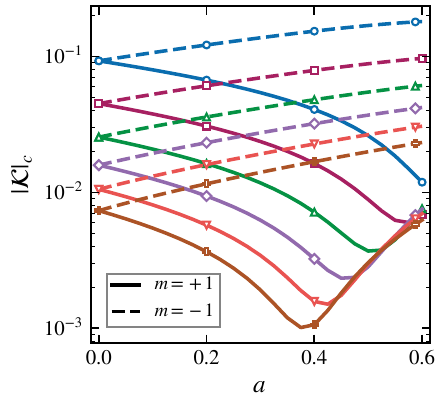}}\hfill
		\subfloat[Logarithmic sector ratio.
		\label{fig:multimode-critical:b}]{%
			\includegraphics[width=0.485\textwidth]{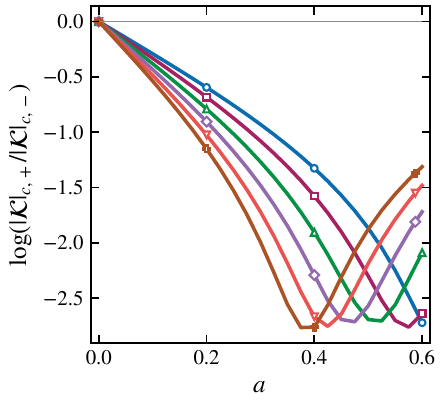}}
		\par\smallskip
		\includegraphics{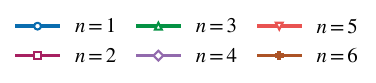}
		\caption{Critical wall reflection magnitudes for six
			branch pairs. Solid and dashed curves denote $m=+1$
			and $m=-1$; colors distinguish branches, with coral
			marking $n=5$. Each pair is degenerate at $a=0$.}
		\label{fig:multimode-critical}
	\end{minipage}
\end{figure*}

All six pairs in Fig.~\ref{fig:multimode-critical} satisfy
$|\Kcal|_{c,+}<|\Kcal|_{c,-}$ for $0<a\leq0.6$, with
pair dependent splitting and turning points. For the representative
pair, this ordering also survives attachment through a smooth
$C^2$ transition layer (Supplemental Table~V),
displacements of $z_w$ and $z_b$
(Supplemental Table~IV), and scans at fixed
circumferential matching radius instead of fixed $z_b$
(Supplemental Fig.~12). The critical values shift
in each comparison. Their ordering is a result for these branches
and geometries, rather than a consequence of rotation-reversal symmetry.

\begin{figure*}[tp]
	\begin{minipage}{6.5in}
		\centering
		\subfloat[Growth and decay rates.
		\label{fig:logK-scaling:a}]{%
			\includegraphics[width=0.485\textwidth]{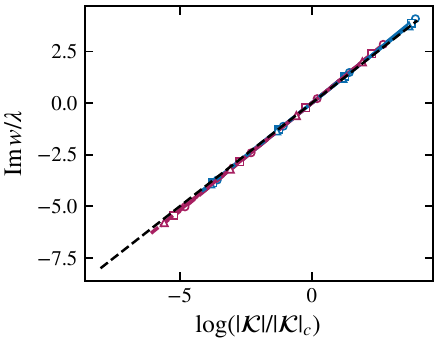}}\hfill
		\subfloat[Oscillation frequency.
		\label{fig:logK-scaling:b}]{%
			\includegraphics[width=0.485\textwidth]{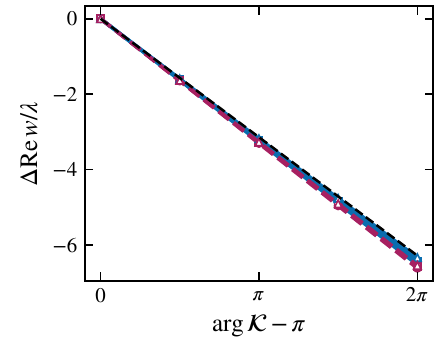}}
		\par\smallskip
		\includegraphics{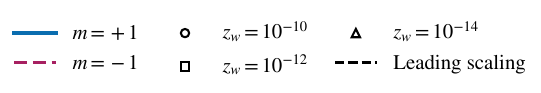}
		\caption{Wall control at $a=0.6$ for
			$z_w=10^{-10},10^{-12},10^{-14}$.
			Frequency changes are scaled by
			$\lambda=(1-a^2)/|\log z_w|$.
			Dashed lines show the leading near-horizon slopes
			$\dd w/\dd\log|\Kcal|=\iu\lambda$ and
			$\dd w/\dd\arg\Kcal=-\lambda$.}
		\label{fig:logK-scaling}
	\end{minipage}
\end{figure*}

The dependence on $\Kcal$ follows from the near-horizon radial
solutions. Let $C_{b,m}(w)$ denote the ratio of outgoing to ingoing
horizon-basis coefficients of the solution that is purely incident
at $x_b$ [Eq.~(A13)]. To leading order in
$z_w$, the wall condition becomes
\begin{equation}
\begin{aligned}
\Kcal&\simeq C_{b,m}(w)e^{-\iu T_{\rm nh}(w-aq)},\\
T_{\rm nh}&=\frac{|\log z_w|}{1-a^2}=\lambda^{-1}.
\end{aligned}
\label{eq:near-horizon-delay}
\end{equation}
Its modulus gives
$T_{\rm nh}\operatorname{Im}w\simeq
\log|\Kcal|-\log|C_{b,m}(w)|$, while its phase gives
$T_{\rm nh}(\operatorname{Re}w-aq)+\arg\Kcal
\simeq\arg C_{b,m}(w)+2\pi n_{\rm ph}$,
with $n_{\rm ph}\in\mathbb Z$ fixed along a local phase branch.
At a real-frequency zero, the modulus equation reduces to the
critical-coupling condition $|\Kcal|_c\simeq|C_{b,m}(w_c)|$.
It is the exterior coefficient, together with the phase condition,
that determines which wall reflection magnitude gives zero reflection.

For a regular simple zero with $\Kcal\ne0$, implicit differentiation
gives
\begin{equation}
\begin{aligned}
\frac{\dd w}{\dd\log\Kcal}
&=-\Kcal\frac{D_K}{D_w}\\
&=-\bigl[\iu T_{\rm nh}-\partial_w\log C_{b,m}
+O(z_w)\bigr]^{-1}.
\end{aligned}
\label{eq:logK-leading-law}
\end{equation}
Here $D_K=\partial_{\Kcal}D$. The logarithm is taken on a local analytic branch; the exact
first line also retains the finite-wall corrections.
When $|\partial_w\log C_{b,m}|\ll T_{\rm nh}$, this yields
$\dd w\simeq\iu\lambda\,\dd\log|\Kcal|
-\lambda\,\dd\arg\Kcal$.
Thus $|\Kcal|$ primarily changes the growth or decay rate,
and $\arg\Kcal$ changes the oscillation frequency, on the same
scale $\lambda$.
Figure~\ref{fig:logK-scaling} shows these slopes and the
sector-dependent deviations due to $\partial_w\log C_{b,m}$.
The delay scale $T_{\rm nh}$ is even in $a$; the directional
dependence remains in both $aq$ and $C_{b,m}$.
The propagation phase is related to those entering echo transfer
functions \cite{Mark:2017echo,Maggio:2019echo}, although
$T_{\rm nh}$ is not a measured echo spacing.

\FloatBarrier

\section{Wave-packet response and energy balance}
\label{sec:time-domain}

\begin{figure*}[tp]
	\begin{minipage}{6.5in}
		\centering
		\subfloat[Reflectance, $m=+1$.\label{fig:multimode-reflection:a}]{%
			\includegraphics[width=0.485\textwidth]{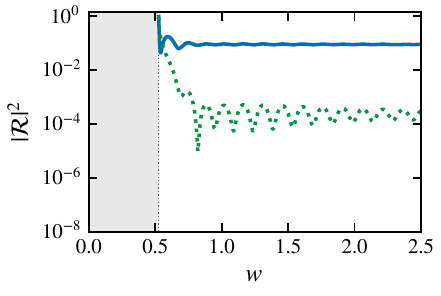}}\hfill
		\subfloat[Reflectance, $m=-1$.\label{fig:multimode-reflection:b}]{%
			\includegraphics[width=0.485\textwidth]{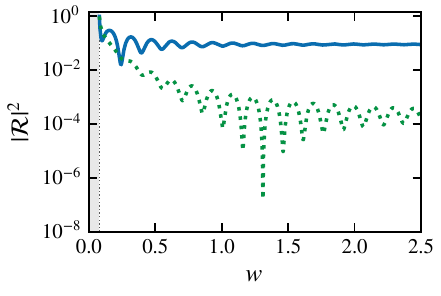}}
		\par\smallskip
		\subfloat[Incident spectrum, $m=+1$.\label{fig:multimode-reflection:input-plus}]{%
			\includegraphics[width=0.485\textwidth]{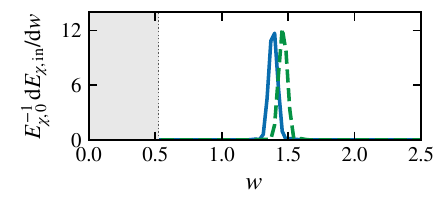}}\hfill
		\subfloat[Incident spectrum, $m=-1$.\label{fig:multimode-reflection:input-minus}]{%
			\includegraphics[width=0.485\textwidth]{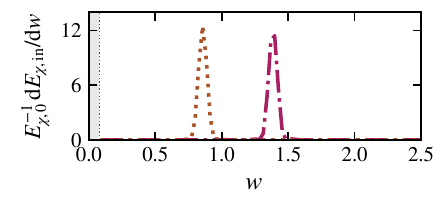}}
		\par\smallskip
		\subfloat[Wall dependence, $m=+1$.\label{fig:multimode-reflection:c}]{%
			\includegraphics[width=0.485\textwidth]{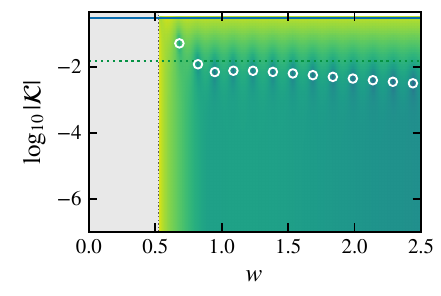}}\hfill
		\subfloat[Wall dependence, $m=-1$.\label{fig:multimode-reflection:d}]{%
			\includegraphics[width=0.485\textwidth]{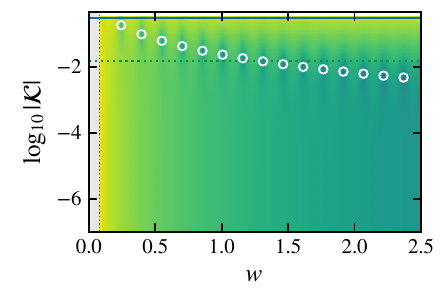}}
		\par\smallskip
		\includegraphics{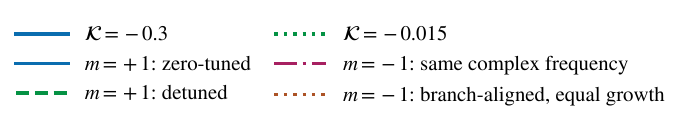}
		\par\smallskip
		\includegraphics[width=0.70\textwidth]{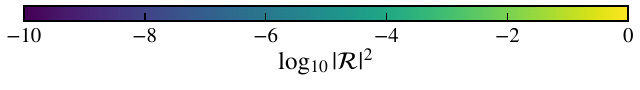}
		\caption{Real-frequency reflectance and incident spectra at
			$a=0.6$. The upper panels show $|\Rcal|^2$ for
			$\Kcal=-0.3,-0.015$; the middle panels show the
			energy-normalized incident spectra of four packets evolved
			at $\Kcal=-0.015$. The lower maps show wall dependence,
			with the two settings marked. Gray regions are
			nonpropagating; white circles mark refined real-frequency
			zeros. The map floor is $10^{-10}$.}
		\label{fig:multimode-reflection}
	\end{minipage}
\end{figure*}

A truncated packet has a finite bandwidth on the real axis, so
its reflection depends on $|\Rcal(\omega)|^2$ across that band
and not on the complex zero alone
\cite{Baranov:2017,Tuncer:2025dnp,Wu:2026vam}.
Figure~\ref{fig:multimode-reflection} shows the reflectance together
with the incident spectra of the four packets defined below.
We retain propagating components with $k_b>0$ and $s_w>0$
(Supplement~C.3).

At the observation point, incident and outgoing amplitudes are
separated by Eq.~(A4), including during
temporal overlap. For
$\chi=\partial_t+\Omega_H\partial_\varphi$, the incident energy
and predicted full-record reflected energy are
\begin{align}
	E_{\chi,{\rm in}}
	&=\int_{\rm prop}\frac{\dd\omega}{2\pi}
	(\omega-m\Omega_H)k_b
	|\widetilde A_{\rm in}(\omega)|^2,
	\label{eq:time-incident-energy}\\
	\frac{E^{\rm pred}_{\chi,{\rm ret}}}{E_{\chi,0}}
	&=\frac{1}{E_{\chi,0}}
	\int_{\rm prop}\frac{\dd\omega}{2\pi}
	(\omega-m\Omega_H)k_b\nonumber\\
	&\qquad\times|\Rcal(\omega)|^2
	|\widetilde A_{\rm in}(\omega)|^2.
	\label{eq:time-predicted-return-fraction}
\end{align}
Here $E_{\chi,0}$ is the initial total energy on
$[x_w,x_{\max}]$. All retained frequencies satisfy
$\omega-m\Omega_H>0$, giving a positive spectral weight.
The plotted density obeys
$\dd E_{\chi,{\rm in}}/\dd w
=(r_+/L^2)\dd E_{\chi,{\rm in}}/\dd\omega$.
The measured full-record reflected energy follows by replacing
$\widetilde A_{\rm in}$ with $\widetilde A_{\rm out}$ in
Eq.~\eqref{eq:time-incident-energy}, where
$\widetilde A_{\rm out}
=e^{2\iu k_b(x_s-x_b)}\Rcal\widetilde A_{\rm in}$.
The propagation phase has unit modulus on the retained band.

We compare four packets with the same initial envelope and
truncation at $\Kcal=-0.015$: an $m=+1$ packet tuned to the
growing $n=5$ zero $w_+$; an $m=+1$ packet with a detuned
oscillation frequency and the same growth rate; an $m=-1$
packet at $w_+$; and an $m=-1$ packet with the oscillation
frequency of its own $n=5$ zero but the same positive growth
rate. The last packet is not tuned to that zero, which requires
a decaying wave. We use each packet's measured incident spectrum
and its group-velocity-dependent incidence interval
(Supplement~C.3).
Tables~\ref{tab:selective-time-controls} and
\ref{tab:selective-time-full-record} report the reflected energy
during incidence and over the full record, including the response
to truncation, respectively.

\label{subsec:time-selective-pulses}

\begin{table}[t]
	\centering
	\small
	\caption{Incident and reflected energy during incidence,
		normalized by $E_{\chi,0}$, at $\Kcal=-0.015$.
		The interval runs from the packet front reaching the wall
		until before the truncated packet center returns;
		its endpoints depend on group velocity.}
	\label{tab:selective-time-controls}
	\begingroup\GridTableSetup
	\setlength{\tabcolsep}{3.5pt}
	\begin{tabular}{|c|c|c|}
		\hline
		Packet & $E_{\chi,{\rm in}}/E_{\chi,0}$
		& \GridCell{Reflected energy\\during incidence} \\
		\hline
		$m=+1$, zero tuned & $0.999971$ & $3.911\times10^{-5}$ \\
		\hline
		$m=+1$, detuned & $0.999978$ & $1.750\times10^{-4}$ \\
		\hline
		\GridCell{$m=-1$,\\same complex frequency}
		& $0.999994$ & $2.959\times10^{-4}$ \\
		\hline
		\GridCell{$m=-1$,\\branch aligned\\same growth rate}
		& $0.999989$ & $1.828\times10^{-4}$ \\
		\hline
	\end{tabular}
	\endgroup
\end{table}

\begin{table}[t]
	\centering
	\small
	\caption{Full-record reflected energy and spectral predictions
		at $\Kcal=-0.015$, normalized by $E_{\chi,0}$.
		The record includes the response after truncation.}
	\label{tab:selective-time-full-record}
	\begingroup\GridTableSetup
	\setlength{\tabcolsep}{3.5pt}
	\begin{tabular}{|c|c|c|}
		\hline
		Packet & \GridCell{Full-record\\reflection}
		& \GridCell{Spectral\\prediction} \\
		\hline
		$m=+1$, zero tuned
		& $1.829\times10^{-4}$ & $1.829\times10^{-4}$ \\
		\hline
		$m=+1$, detuned
		& $3.534\times10^{-4}$ & $3.546\times10^{-4}$ \\
		\hline
		\GridCell{$m=-1$,\\same complex frequency}
		& $5.662\times10^{-4}$ & $5.685\times10^{-4}$ \\
		\hline
		\GridCell{$m=-1$,\\branch aligned\\same growth rate}
		& $7.560\times10^{-4}$ & $7.590\times10^{-4}$ \\
		\hline
	\end{tabular}
	\endgroup
\end{table}

The zero-tuned packet reflects least by either measure
(Tables~\ref{tab:selective-time-controls} and
\ref{tab:selective-time-full-record}). Over the full record,
it reflects $1.83\times10^{-4}$ of its initial energy, compared
with $(3.53$--$7.56)\times10^{-4}$ for the three controls,
a reduction by factors of $1.9$--$4.1$.
Equation~\eqref{eq:time-predicted-return-fraction}, evaluated
with each measured incident spectrum, reproduces all four
full-record values to within $0.42\%$.
For the counterrotating controls, moving the oscillation
frequency onto the $m=-1$ branch lowers the reflected energy
during incidence from $2.96\times10^{-4}$ to
$1.83\times10^{-4}$ of the initial energy, but raises the
full-record fraction from $5.66\times10^{-4}$ to
$7.56\times10^{-4}$. This packet retains a growing envelope,
whereas the zero requires decay; the comparison tests a change
of oscillation frequency at fixed growth rate.

\begin{figure*}[tp]
	\centering
	\begin{minipage}{6.5in}
		\centering
		\subfloat[Cumulative reflected energy.
		\label{fig:selective-time-dynamics:a}]{%
			\includegraphics[width=0.425\textwidth]{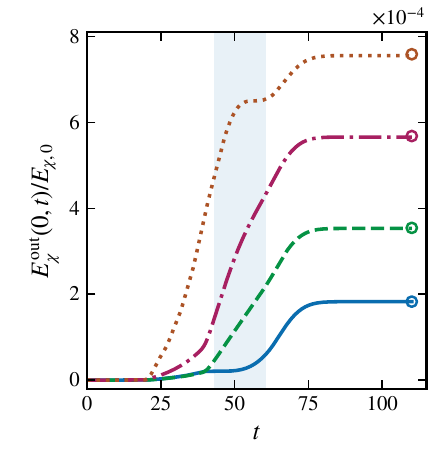}}\hfill
		\subfloat[Regional energy and wall loss.
		\label{fig:selective-time-dynamics:b}]{%
			\includegraphics[width=0.425\textwidth]{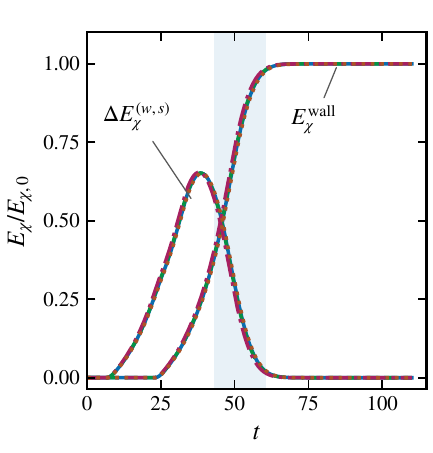}}
		\par\smallskip
		\includegraphics{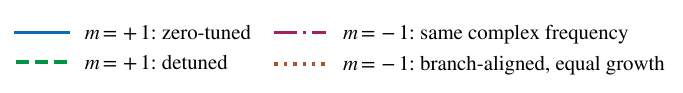}
		\caption{Energy evolution for four packets at
			$\Kcal=-0.015$, normalized by their respective $E_{\chi,0}$.
			In (a), open endpoints show spectral predictions for the
			full-record reflected energy. Panel (b) shows regional energy changes
			$\Delta E_\chi^{(w,s)}$ and cumulative wall loss
			$E_\chi^{\rm wall}$; the region includes part of the exterior
			channel. Shading marks the zero-tuned packet's incidence
			interval.}
		\label{fig:selective-time-dynamics}
	\end{minipage}
	\par\medskip
	\begin{minipage}{6.5in}
		\centering
		\subfloat[Trajectories in the complex-frequency plane.\label{fig:multimode-wall:a}]{%
			\includegraphics[width=0.425\textwidth]{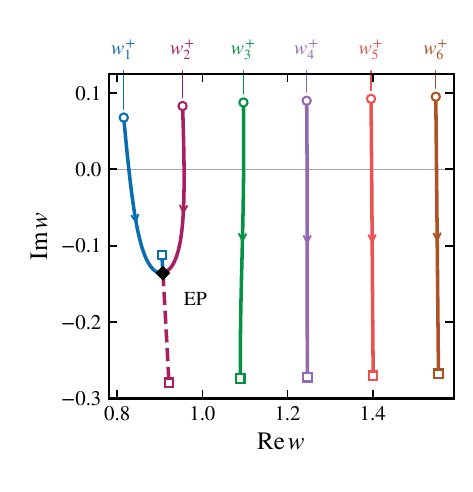}}\hfill
		\subfloat[Coalescing pair: imaginary parts.\label{fig:multimode-wall:d}]{%
			\includegraphics[width=0.425\textwidth]{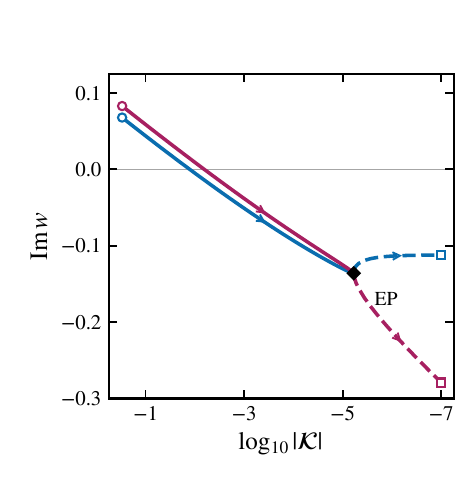}}
		\par\smallskip
		\includegraphics{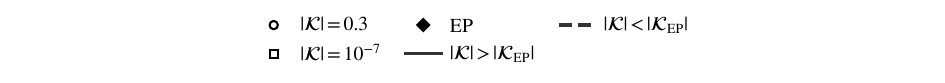}
\caption{Reflection-zero trajectories at $a=a_{\rm EP}$ and $m=+1$,
with $|\Kcal|$ decreasing from $0.3$ to $10^{-7}$ along the negative real axis.
Panel~\protect\subref{fig:multimode-wall:a} shows six branches;
panel~\protect\subref{fig:multimode-wall:d} shows the imaginary parts of the
$n=1,2$ pair. The black diamond marks the EP.
Solid and dashed segments lie at $|\Kcal|>|\Kcal_{\rm EP}|$ and
$|\Kcal|<|\Kcal_{\rm EP}|$, respectively.}
		\label{fig:multimode-wall}
	\end{minipage}
\end{figure*}

The energy between the wall and the observation point is
$E_\chi^{(w,s)}=\int_{x_w}^{x_s}\rho_\chi\dd x$, where
$\rho_\chi=\rho_E-\Omega_H\rho_J$ and
$F_\chi=F_E-\Omega_HF_J$.
The integration interval includes the exterior segment $[x_b,x_s]$.
With positive flux toward increasing $x$, conservation gives
\begin{equation}
\begin{aligned}
\frac{\dd E_\chi^{(w,s)}}{\dd t}
&=F_\chi(x_w,t)-F_\chi(x_s,t),\\
E_\chi^{\rm wall}(t)
&=-\int_0^tF_\chi(x_w,t')\dd t'.
\end{aligned}
\label{eq:time-control-volume-balance}
\end{equation}
Figure~\ref{fig:selective-time-dynamics}\subref{fig:selective-time-dynamics:b}
plots $\Delta E_\chi^{(w,s)}=E_\chi^{(w,s)}(t)-E_\chi^{(w,s)}(0)$
and $E_\chi^{\rm wall}$.
The four histories are nearly indistinguishable: the regional
energy peaks near $0.65E_{\chi,0}$, and more than $99.9\%$ of
$E_{\chi,0}$ has crossed the wall by the end of each record.
For these four packets the weakly reflecting wall absorbs
nearly all the energy, so tuning to the zero changes the
reflected remainder of order $10^{-4}E_{\chi,0}$ while barely
changing wall loss or peak regional energy.

The wall is passive in the sense
$F_E-\Omega_wF_J\leq0$, with $\Omega_w=-N_w^\varphi$
[Eq.~(C5)]. This differs from the
$\chi$ flux because $\Omega_w\ne\Omega_H$; the cumulative
$\chi$ loss is nevertheless nonnegative for all four packets.
Spatial refinement changes the full-record reflected energies
by at most $2.6\%$, with their ordering unchanged
(Supplement~C.3).
\label{subsec:time-current-accounting}

\FloatBarrier

\section{Exceptional point of the reflection zeros}
\label{sec:reflection-zero-ep}

\subsection{Coalescence and local structure}

Two corotating reflection zeros, $n=1,2$, coalesce at
$a_{\rm EP}\simeq0.592$, $\Kcal_{\rm EP}\simeq-5.97\times10^{-6}$,
and $w_{\rm EP}\simeq0.9074-0.1359\iu$.
The negative imaginary part requires an exponentially decaying incident wave;
the growing wave packets of Sec.~\ref{sec:time-domain} use the distinct $n=5$ pair.
As $|\Kcal|$ decreases at $a=a_{\rm EP}$, the higher-frequency zeros in
Fig.~\ref{fig:multimode-wall} descend with shifts in $\operatorname{Re}w$ smaller than
$0.008$. The low-frequency pair approaches, coalesces, and separates primarily
in $\operatorname{Im}w$: one decay rate decreases while the other increases.
Beyond the coalescence, the plotted $n=1$ continuation denotes the less-damped
member and $n=2$ the more-damped member. This convention follows the two local
square-root signs; the straight real-$\Kcal$ path does not uniquely identify
an individual branch through the EP.

\begin{figure*}[!t]
	\begin{minipage}{6.5in}
		\centering
		\subfloat[Spin determined by the real wall condition.\label{fig:ep-length-control:a}]{%
			\includegraphics[width=0.485\textwidth]{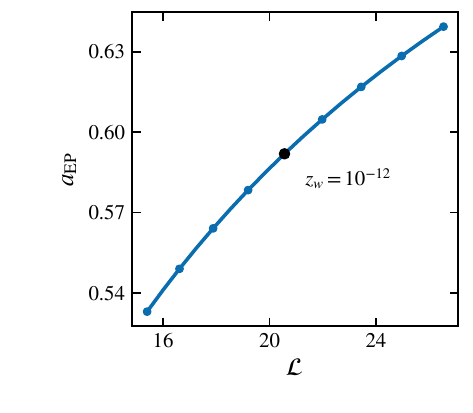}}\hfill
		\subfloat[Reflection magnitude under two constraints.\label{fig:ep-length-control:b}]{%
			\includegraphics[width=0.485\textwidth]{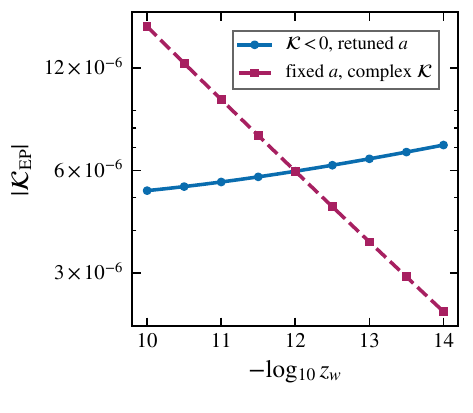}}
		\caption{
			Dependence of the $m=+1$ reflection-zero EP on cavity length,
			with $r_b/r_+\simeq1.1618$ and the remaining field parameters
			fixed as in Table~\ref{tab:ttm-parameters}.
			Blue circles give the negative real $\Kcal$ path,
			with both the spin and the reflection magnitude retuned to maintain the double root.
			The cavity length is
			$\mathcal L=(r_+/L^2)(x_b-x_w)$.
			The black point marks the reference wall position
			$z_w=10^{-12}$.
			Magenta squares correspond to $a=a_{\rm EP}$
			with complex $\Kcal$, included as a mathematical comparison.
			Markers are EP solutions and lines connect neighboring wall positions.
		}
		\label{fig:ep-length-control}
	\end{minipage}
\end{figure*}

The coalescence occurs where the ordinary frequency-control relation of
Sec.~\ref{sec:critical-coupling} ceases to be invertible.
Since $D=D_1(\Kcal-\Kcal_{\rm req})$ with $D_1\ne0$, a double zero at fixed
$\Kcal$ is a stationary point of $\Kcal_{\rm req}(w)$.
In the near-horizon approximation,
$\partial_w\log C_{b,+1}\simeq\iu T_{\rm nh}$ at this point:
the exterior term is now as large as the near-horizon term in
Eq.~\eqref{eq:logK-leading-law}.
Near a simple zero $w_{\rm ext}$ of $C_{b,+1}$, write
$C_{b,+1}\simeq c(w-w_{\rm ext})$, where $c=C'_{b,+1}(w_{\rm ext})$.
The stationary point of
$\Kcal_{\rm req}\simeq c(w-w_{\rm ext})\exp[-\iu T_{\rm nh}(w-aq)]$
then gives
\begin{equation}
 \begin{aligned}
 w_{\rm EP}^{(0)}&=w_{\rm ext}-\frac{\iu}{T_{\rm nh}},\\
 |\Kcal_{\rm EP}^{(0)}|&=\frac{|c|}{\mathrm e T_{\rm nh}}
 \exp[T_{\rm nh}\operatorname{Im}w_{\rm ext}].
 \end{aligned}
 \label{eq:ep-length-leading}
\end{equation}
At $a=a_{\rm EP}$, $w_{\rm ext}\simeq0.90492-0.11201\iu$ and
$T_{\rm nh}\simeq42.535$, giving
$w_{\rm EP}^{(0)}\simeq0.90492-0.13552\iu$.
The negative $\operatorname{Im}w_{\rm ext}$ exponentially suppresses the required
$|\Kcal|$. Along the cavity-length family below, the leading prediction for the complex
$\Kcal_{\rm EP}$ differs from the numerical value by $10$--$11\%$;
including the curvature of $C_{b,+1}$ reduces this to $1.2$--$1.4\%$
(Supplement~B.1).
In this leading approximation, $w_{\rm ext}$ is also the zero for
$\Kcal=0$. The finite-position Robin condition adds an $O(z_w)$ correction,
so the exact zero at $\Kcal=0$ is instead defined by $D_0=0$.

At the numerical EP,
\begin{equation}
 D=D_w=0,\qquad D_{ww}\ne0,\qquad D_K\ne0.
 \label{eq:ep-defect-condition}
\end{equation}
The purely incident condition at $x_b$ fixes the radial solution up to
normalization, so the double root has a single eigenfunction.
The frequency-dependent boundary-value problem has a Jordan chain of length
two, including the derivatives of both boundary conditions
\cite{Guttel:2017nep}; Supplement~B gives the proof.
Expanding at fixed $z_w,z_b$ yields
\begin{equation}
 \begin{aligned}
 (\delta w)^2&\simeq\beta_a\delta a+\beta_K\delta\Kcal,\\
 \beta_p&=-\left.\frac{2D_p}{D_{ww}}\right|_{\rm EP}.
 \end{aligned}
 \label{eq:ep-local-unfolding}
\end{equation}
A circuit around the EP in the real $(a,\Kcal)$ plane exchanges the two zeros;
two circuits restore them (Supplemental Fig.~14).
For a simple zero, the sensitivity
$\mathcal S_K=\dd w/\dd\log|\Kcal|=-\Kcal D_K/D_w$
in Eq.~\eqref{eq:logK-leading-law} consequently diverges as
$|\mathcal S_K|\simeq C|\varepsilon|^{-1/2}$, where
$\varepsilon=(\Kcal-\Kcal_{\rm EP})/|\Kcal_{\rm EP}|$ at $a=a_{\rm EP}$ and
$C=\tfrac12\sqrt{|\Kcal_{\rm EP}|\,|\beta_K|}$.
The static spin sweep in Supplemental Fig.~13 passes through the same EP.

\subsection{Continuation in cavity length}

The double-root conditions are four real equations for the complex frequency
and the controls. At fixed wall position, the real spin and real negative
$\Kcal$ provide the two controls needed for an isolated EP.
Varying the wall position traces the line in Fig.~\ref{fig:ep-length-control},
with the circumferential matching radius held fixed.
For $\mathcal L=(r_+/L^2)(x_b-x_w)=15.40$--$26.54$,
$a_{\rm EP}$ increases from $0.533$ to $0.639$ and $|\Kcal_{\rm EP}|$ increases
by about $36\%$.
If the spin is instead fixed at $a_{\rm EP}$ and $\Kcal$ is complex,
$|\Kcal|$ and $\arg\Kcal$ are independent controls; over the same range of
wall positions the required magnitude decreases from $1.59\times10^{-5}$
to $2.31\times10^{-6}$.

The difference follows from the phase of the same near-horizon map.
Keeping $\Kcal$ negative and real forces the spin to change with the wall
position through Eq.~(B10).
Along this path, $|C'_{b,+1}(w_{\rm ext})|$ increases enough to outweigh the
length and damping terms in Eq.~(B11).
The independently adjustable phase of complex $\Kcal$ removes this constraint.
The comparison covers the stated finite range of wall positions.
The EP also persists at shifted parameters when a smooth transition layer
connects the BTZ region to the exterior channel
(Supplemental Table~VI).

\subsection{Response to a localized source}

\begin{figure*}[tp]
	\begin{minipage}{6.5in}
		\centering
		\subfloat[$\varepsilon=-0.03$.\label{fig:ep-complex-response:a}]{%
			\includegraphics[width=.32\textwidth]{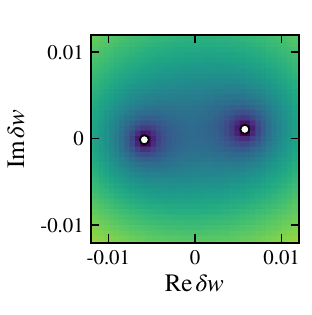}}\hfill
		\subfloat[$\varepsilon=0$.\label{fig:ep-complex-response:b}]{%
			\includegraphics[width=.32\textwidth]{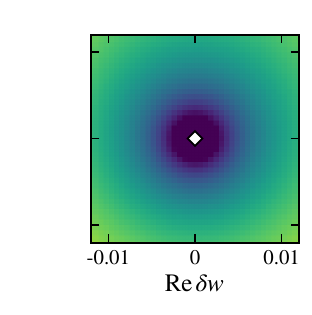}}\hfill
		\subfloat[$\varepsilon=+0.03$.\label{fig:ep-complex-response:c}]{%
			\includegraphics[width=.32\textwidth]{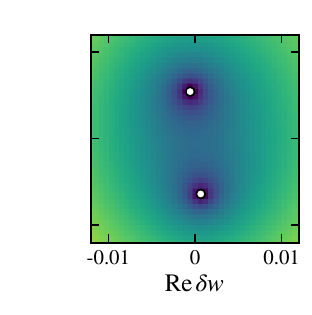}}
		\par\smallskip
		\includegraphics[width=3.8in]{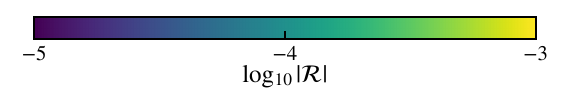}
		\par\smallskip
		\subfloat[\label{fig:ep-complex-response:d}]{%
			\includegraphics[width=.32\textwidth]{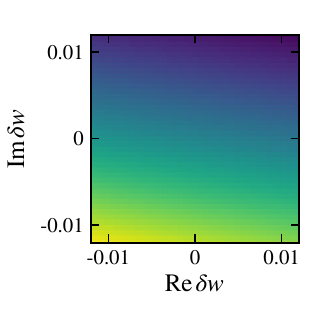}}\hfill
		\subfloat[\label{fig:ep-complex-response:e}]{%
			\includegraphics[width=.32\textwidth]{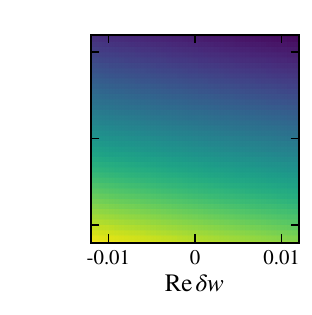}}\hfill
		\subfloat[\label{fig:ep-complex-response:f}]{%
			\includegraphics[width=.32\textwidth]{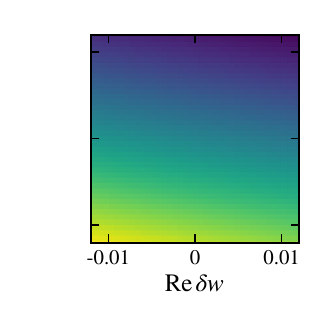}}
		\par\smallskip
		\includegraphics[width=3.8in]{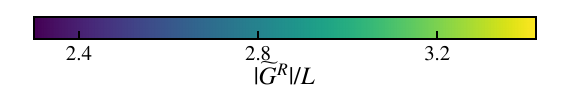}
		\caption{Reflection zeros and the analytically continued retarded
			Green function near the $m=+1$ EP, with
			$\delta w=w-w_{\rm EP}$, $z'=0.05$, and $x_s=5$,
			at the fixed reference EP spin and geometry.
			The upper row shows $\log_{10}|\Rcal|$, with a display floor
			of $10^{-5}$ for the reflection amplitude.
			Circles mark simple zeros,
			and diamonds mark double zeros.
			The lower row shows $|\widetilde G_m^R|/L$
			using a common color scale and the same column parameters as the upper row.
			All panels cover the same window in the lower half
			of the complex-frequency plane.}
		\label{fig:ep-complex-response}
	\end{minipage}
\end{figure*}

For a reduced unit source at $x'$ and an observation point in the exterior
channel, the retarded Green function is
\begin{equation}
 \widetilde G_m^R(x_s,x';\omega)
 =-(1+\Kcal)\frac{u_w(x',\omega)e^{\iu k_b(x_s-x_b)}}{E(\omega)},
 \label{eq:ep-local-source-green}
\end{equation}
where $u_w$ satisfies Eq.~\eqref{eq:ttm-wall-impedance} and
$u_w(x_w)=1$.
Its denominator is the outgoing-wave function $E$ in $\Rcal=-D/E$
\cite{Mark:2017echo}.
At fixed geometry and frequency, both $(1+\Kcal)u_w$ and $E$ are affine in
$\Kcal$, so the source response is analytic in $\Kcal$ wherever $E\ne0$.
With this normalization, $|E_{\rm EP}|\simeq0.153$ and the sampled minimum
of $|E|$ in Fig.~\ref{fig:ep-complex-response} is $0.118$.
The reflection zeros merge in the upper row, while the lower row shows the
Green function in the same lower-half-plane window.

\begin{figure*}[tp]
	\begin{minipage}{6.5in}
		\centering
		\subfloat[Amplitude spectrum, $m=+1$.\label{fig:green-broad-spectra:a}]{%
			\includegraphics[width=.485\textwidth]{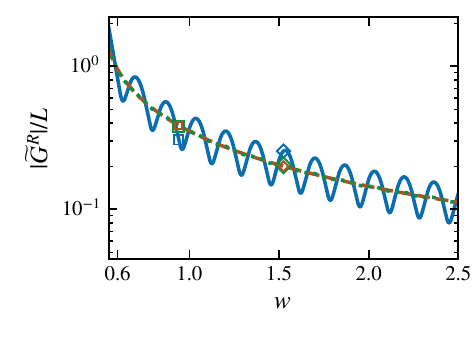}}\hfill
		\subfloat[Amplitude spectrum, $m=-1$.\label{fig:green-broad-spectra:b}]{%
			\includegraphics[width=.485\textwidth]{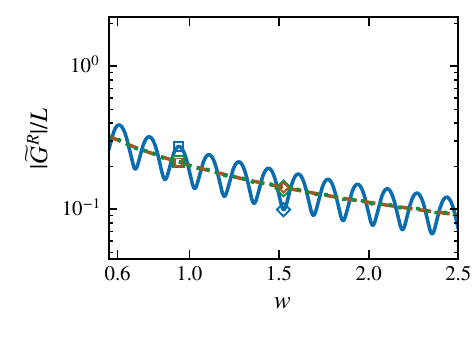}}
		\par\medskip
		\subfloat[Complex response, $m=+1$.\label{fig:green-broad-spectra:c}]{%
			\includegraphics[width=.485\textwidth]{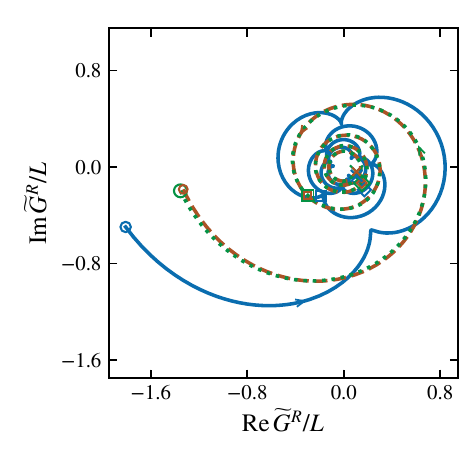}}\hfill
		\subfloat[Complex response, $m=-1$.\label{fig:green-broad-spectra:d}]{%
			\includegraphics[width=.485\textwidth]{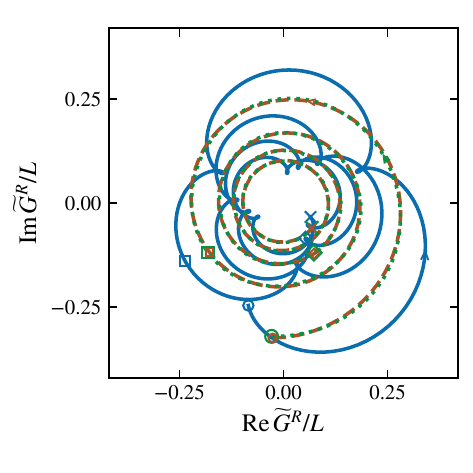}}
		\par\smallskip
		\includegraphics[width=6.5in]{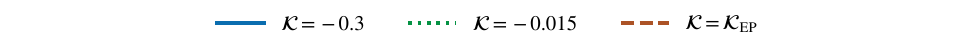}
		\par\smallskip
		\includegraphics[width=6.5in]{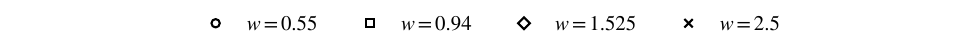}
		\caption{Response to a localized source at
			$a=a_{\rm EP}$, $z'=0.05$, and $x_s=5$.
			The upper panels show amplitude spectra at real frequencies,
			while the lower panels show trajectories of the complex Green function
			as $w$ increases from $0.55$ to $2.5$.
			Open circles and crosses mark the two ends of the frequency interval;
			squares and diamonds identify the same two sampled frequencies in all panels,
			and open arrowheads indicate the direction of increasing $w$.
			All panels use the same three wall settings and normalization
			to a reduced unit source, but the axis ranges differ between the two sectors.
			In both sectors, $\Kcal_{\rm EP}$ denotes the wall coefficient
			corresponding to the $m=+1$ EP.}
		\label{fig:green-broad-spectra}
	\end{minipage}
\end{figure*}

For $z'=0.05,0.20$ and $x_s=5$, the relative band norm
$\mathcal N_G$ of Eq.~(B12), integrated over
$0.55\leq w\leq2.5$, scales linearly with $|\varepsilon|$ for
$10^{-8}\leq|\varepsilon|\leq10^{-3}$.
Over that interval the half splitting of the zeros scales as
$|\varepsilon|^{0.50}$.
Thus the square-root displacement of the zeros does not describe the
perturbation of the field generated by a fixed source.
At larger changes in $\Kcal$, Fig.~\ref{fig:green-broad-spectra} compares
both the amplitude and phase of this field.
For $m=+1$ and $z'=0.05$, the responses at $\Kcal=-0.3$ and $-0.015$
differ from the $\Kcal_{\rm EP}$ response by about $32\%$ and $1.6\%$,
respectively, in the same band norm.
The oscillations and complex plane loops are more pronounced at
$\Kcal=-0.3$; the two smaller magnitudes give nearly coincident curves.
The source normalization and independent propagation check are given in
Supplement~B.2.

\section{Conclusions}
\label{sec:scope}

This paper studies the scattering of a scalar field by a partially reflecting surface located near the horizon in the rotating BTZ spacetime, and determines the conditions for the existence of reflection zeros, the effects of rotation, and their exceptional point (EP) structure. We find that, even when the near horizon wall has nonzero reflectivity, the scattering system as a whole can achieve zero reflection for incident waves of specific complex frequencies. This shows that reflection zeros are not determined solely by the local reflective properties of the wall; rather, they arise from the combined action of the near horizon boundary condition, wave propagation, and external scattering.

Rotation significantly modifies the distribution of reflection zeros and their time-domain behaviour. For the paired branches considered here, the corotating and counterrotating modes possess different critical wall reflection coefficients, so that there exists a parameter range in which the corotating incident wave required for zero reflection grows with time while the counterrotating incident wave decays with time. A near horizon analysis further reveals how the zero frequencies depend on the logarithmic amplitude and the phase of the wall reflection coefficient. In the time domain, we find that tuning a truncated incident wave packet to a reflection zero further suppresses the residual reflection, but leaves the wall absorption and the energy accumulation in the cavity almost unchanged. This indicates that tuning to a reflection zero mainly controls the reflected fraction in the external scattering channel, rather than the overall energy dissipation of the system.

Another main result of this paper is the finding that two corotating reflection-zero branches can merge at an EP and exhibit a characteristic square-root frequency splitting. Using a near horizon inverse mapping, we determine the location of the EP and the corresponding wall reflection condition. Notably, the merging of reflection zeros does not imply that the poles of the Green's function for a localized source also merge: the latter are still determined by an independent outgoing boundary condition. Therefore, the spectral sensitivity embodied by an EP of reflection zeros cannot be identified directly with an enhancement of the signal excited by a physical source.

These results reveal a class of nontrivial scattering structures arising from the combined action of near horizon reflection and rotation, and they also illustrate the importance of distinguishing scattering zeros from the intrinsic resonances of the system. This paper adopts a finite BTZ interval and prescribed external matching conditions, so the critical reflection coefficients and the specific location of the EP are not universal. It would be worthwhile in the future to extend this analysis to black hole geometries with more realistic external scattering potentials, and to introduce a frequency dependent wall response that respects causality, in order to study how reflection zeros and their EPs affect the signals produced by physical sources, in particular the observable late-time echo structure \cite{Mark:2017echo,Maggio:2019echo,Rosato:2025greybody,Chakraborty:2022membrane}.

%
%
%
%

\begin{acknowledgments}
	This work was supported by the National Natural Science Foundation of China (Grants No. 12547143, 12305076, 12405055 and 12305063
), the Key Project of the Department of Education of Hunan Province (No.
25A0084), the Hunan Provincial Furong Plan Youth Talent Support Program (Grant No. 2026ZRB-QN100).
\end{acknowledgments}

\FloatBarrier
\bibliographystyle{apsrev4-2}
\nocite{DLMF:LambertW,Trefethen:2005,Jaramillo:2020tuu,Gasperin:2021pseudospectrum,Geissler:2026hankel}
\bibliography{references}

\end{document}


\setcounter{figure}{11}
\setcounter{table}{3}
\appendix
\section*{Supplemental Material}

\section{Exact radial solutions and matching expansions}
\label{app:exact-checks}

The radial effective potential is
\begin{equation}
	V_m=N^2\left(\frac{m^2}{r^2}+\mu^2\right)
	+\frac{N^2(N^2)'}{2r}-\frac{N^4}{4r^2},
	\label{eq:ttm-potential}
\end{equation}
where a prime denotes $\dd/\dd r$. At the matching surface $x_b$, the unit incident and outgoing radial solutions satisfy
\begin{equation}
	\begin{aligned}
		(j,j_x)_{x_b}&=(1,-\iu k_b),\\
		(o,o_x)_{x_b}&=(1,+\iu k_b).
	\end{aligned}
	\label{eq:ep-jost-data}
\end{equation}
Defining $\widehat B_w=(1+\Kcal)\mathcal B_w^{(Z)}$, the regular characteristic function depends linearly on the wall reflection coefficient $\Kcal$:
\begin{equation}
	\begin{aligned}
		D&=\widehat B_w[j]=D_0+\Kcal D_1,\\
		D_0&=j_x(x_w)+\iu s_wj(x_w),\\
		D_1&=j_x(x_w)-\iu s_wj(x_w),\qquad E=\widehat B_w[o].
	\end{aligned}
	\label{eq:ep-regular-characteristic}
\end{equation}
At $x_b$, the incident and outgoing amplitudes are obtained by projecting the radial field and its derivative:
\begin{equation}
	\Ain=\frac12\left(\Psi-\frac{\Psi_x}{\iu k_b}\right),\qquad
	\Aout=\frac12\left(\Psi+\frac{\Psi_x}{\iu k_b}\right).
	\label{eq:ttm-port-amplitudes}
\end{equation}
The corresponding amplitudes at the observation point differ only by exterior propagation phases, which have unit modulus in the real-frequency propagation regime.

\subsection{Radial basis solutions and finite-radius inversion}

Let $h_-=(1-\nu)/2$ and define
\begin{equation}
	\begin{aligned}
		\alpha&=-\frac{\iu(w-aq)}{2(1-a^2)},\\
		\mathsf A&=h_- -\frac{\iu(w+q)}{2(1+a)},\\
		\mathsf B&=h_- -\frac{\iu(w-q)}{2(1-a)},\\
		\mathsf C&=1-\frac{\iu(w-aq)}{1-a^2}.
	\end{aligned}
	\label{eq:ttm-hyper-parameters}
\end{equation}
For $R=\Psi/\sqrt r$, the ingoing and outgoing radial solutions at the horizon are
\begin{equation}
	\begin{aligned}
		R_{\rm in}(z)={}&z^\alpha(1-z)^{h_-}\\
		&\times{}_2F_1(\mathsf A,\mathsf B;\mathsf C;z),
	\end{aligned}
	\label{eq:ttm-hyper-in}
\end{equation}
\begin{equation}
	\begin{aligned}
		R_{\rm out}(z)={}&z^{-\alpha}(1-z)^{h_-}\\
		&\times{}_2F_1\bigl(\mathsf A-\mathsf C+1,\mathsf B-\mathsf C+1;\\
		&\hspace{51pt}2-\mathsf C;z\bigr).
	\end{aligned}
	\label{eq:ttm-hyper-out}
\end{equation}
The ingoing and outgoing labels refer to the phases near the horizon. Matching at finite boundaries uses the full hypergeometric functions. Where these functions are regular and their Wronskian is nonzero, they form a local basis.

The inverse radial transformation is $r(z)=[(r_+^2-zr_-^2)/(1-z)]^{1/2}$. Differentiating the exact basis functions and projecting them at the finite boundaries gives the incident and outgoing amplitudes in the exterior channel. Changing the observation point within this channel introduces only additional propagation phases.

The hypergeometric solutions were checked by substitution into the radial equation
and wall condition in multiprecision arithmetic. Replacing the exact wall data
by the leading powers $z^{\pm\alpha}$ changes the wall matching by at most
$2.1\times10^{-13}$ at the two tested zeros.

Equation~\eqref{eq:ep-regular-characteristic} directly gives the wall coefficient required for reflectionless matching:
\begin{equation}
	\Kcal_{{\rm req},m}(w)=-\frac{D_0(w)}{D_1(w)}.
	\label{eq:app-required-reflectivity}
\end{equation}
This expression applies only when $D_1\ne0$. Elsewhere, the regular root condition is imposed using the incident-normalized characteristic function $D$. In the real-frequency propagation regime, critical matching with a passive negative real wall coefficient requires $\operatorname{Im}\Kcal_{{\rm req},m}=0$ and $-1<\Kcal_{{\rm req},m}<0$. The corresponding branches are identified by continuous continuation, and the reflection zeros are verified by directly evaluating $\Rcal$.

\subsection{Small spin expansion and local wall control}
\label{app:logK-control}

For real-frequency crossings at negative real wall coefficients, define the local logarithmic function
\begin{equation}
	H_m(w,a)=\operatorname{Log}[-\Kcal_{{\rm req},m}(w,a)].
	\label{eq:critical-log-map}
\end{equation}
We choose an analytic branch of the logarithm that is real at the crossing. Let $\eta=\log|\Kcal|$. The critical matching conditions are then $\operatorname{Im}H_m=0$ and $\eta=\operatorname{Re}H_m$. For the selected nonrotating branch, the critical values are $w_{c,0}=1.8013360133$, $|\Kcal|_{c,0}=0.010530928812$, and $\eta_{c,0}=-4.5534387505$. All derivatives below are evaluated at this crossing and at $a=0$, with $z_w,z_b$ fixed.

Under reversal of rotation, $(a,m)\mapsto(-a,-m)$, the quantities $N^2,r(z),x(z)$ are even in spin, while $N^\varphi$ is odd. The radial equation and both boundary conditions are therefore invariant. In the exact basis, $\mathsf A$ and $\mathsf B$ are interchanged, while $\alpha$ and $\mathsf C$ remain unchanged. The corresponding analytic branches of the logarithm consequently satisfy $H_m(w,a)=H_{-m}(w,-a)$.

Let $H_w=(\partial_wH_+)_0$ and $H_a=(\partial_aH_+)_0$. Away from channel thresholds and singularities of the representation, the conditions $D_1\ne0$ and $\operatorname{Im}H_w\ne0$ ensure a locally unique transverse critical crossing. A simple complex-frequency reflection zero alone does not guarantee the latter condition. Differentiating the two real matching equations gives
\begin{equation}
	\begin{aligned}
		w_1&=-\frac{\operatorname{Im}H_a}{\operatorname{Im}H_w},\\
		\eta_1&=\operatorname{Re}H_a+\operatorname{Re}H_w\,w_1.
	\end{aligned}
	\label{eq:small-spin-coefficients}
\end{equation}
Thus, for $m=\pm1$, we obtain $w_{c,m}=w_{c,0}+maw_1+O(a^2)$ and $\eta_{c,m}=\eta_{c,0}+ma\eta_1+O(a^2)$. Local uniqueness and rotation reversal symmetry further imply $(w_c,\eta_c)_+(a)=(w_c,\eta_c)_-(-a)$, so the differences between the sectors have the odd power expansions in Eq.~\eqref{eq:small-spin-splitting}. Symmetry neither requires $w_1,\eta_1$ to be nonzero nor determines their signs. The coefficients reported in the main text include the full exterior matching and finite wall corrections.

For $a\leq0.02$, the relative deviations of $\Delta w_c$ and $\Delta \eta_c$
from their linear predictions are below $4.3\times10^{-5}$ and
$6.5\times10^{-4}$; at $a=0.1$ they reach $1.12\times10^{-3}$ and
$1.66\times10^{-2}$. The remainders divided by $a^3$ approach finite values,
consistent with the odd-power expansion. The finite-spin curves use the full
radial and exterior matching conditions.

For the representative $n=5$ branch pair at $a=0.6$, the critical crossings with negative real wall coefficients are
$(|\Kcal|_{c,+},w_{c,+})=(0.00698027,1.384869)$ and
$(|\Kcal|_{c,-},w_{c,-})=(0.03069898,0.854611)$.
The corresponding small-spin coefficients are $w_1\simeq0.48973$ and $\eta_1\simeq-2.38666$, including the full radial matching corrections. At the common wall coefficient $\Kcal=-0.015$ used in the wave-packet comparison, continuation gives $w_+=1.384580+0.018258\iu$ and $w_-=0.855423-0.017301\iu$. The ideal reflectionless inputs in the two sectors therefore require exponentially growing and decaying temporal envelopes, respectively.

For a simple reflection zero at $\Kcal\ne0$, local holomorphy gives $\partial_\phi w=\iu\partial_\eta w$, where $\phi=\arg\Kcal$. At $a=0.6$ and $\Kcal=-0.3$, the maximum finite-difference Cauchy--Riemann residual among the six tested $(z_w,m)$ branches is $3.6\times10^{-11}$. The exact derivative in Eq.~\eqref{eq:logK-leading-law} retains the exterior matching corrections to the common leading scale $\lambda$.

Continuous variation of the wall reflection phase also extends the local control to neighboring reflection zeros. At $|\Kcal|=0.3$, three successive $2\pi$ phase cycles connect
\ensuremath{w_{5}^{+}}, \ensuremath{w_{4}^{+}}, \ensuremath{w_{3}^{+}}, and \ensuremath{w_{2}^{+}}
in the $m=+1$ sector, and
\ensuremath{w_{5}^{-}}, \ensuremath{w_{4}^{-}}, \ensuremath{w_{3}^{-}}, and \ensuremath{w_{2}^{-}}
in the $m=-1$ sector.
At the end of each cycle, the physical wall coefficient returns to its original value, but continuous continuation of the inverse matching map selects another reflection zero.

\subsection{Exterior completion and boundary locations}
\label{subsec:smooth-completion}

\begin{table*}[t]
\centering
\caption{Critical wall reflection magnitudes for the selected paired branches at $a=0.6$ in the constant-coefficient exterior model. All nine sampled wall and matching-surface configurations retain $|\Kcal|_{c,+}<|\Kcal|_{c,-}$, while the critical values and their ratio depend on the boundary locations.}
\label{tab:boundary-position-matching}
\begingroup\GridTableSetup
\begin{tabular}{|c|c|c|c|c|}
\hline
$z_w$ & $z_b$ & $|\Kcal|_{c,+}$ & $|\Kcal|_{c,-}$ & $|\Kcal|_{c,-}/|\Kcal|_{c,+}$ \\
\hline
$10^{-10}$ & $0.25$ & $0.004792$ & $0.011774$ & $2.4572$ \\
\hline
$10^{-10}$ & $0.35$ & $0.005908$ & $0.020022$ & $3.3893$ \\
\hline
$10^{-10}$ & $0.45$ & $0.006248$ & $0.031837$ & $5.0956$ \\
\hline
$10^{-12}$ & $0.25$ & $0.005925$ & $0.017873$ & $3.0166$ \\
\hline
$10^{-12}$ & $0.35$ & $0.006980$ & $0.030699$ & $4.3980$ \\
\hline
$10^{-12}$ & $0.45$ & $0.007139$ & $0.049663$ & $6.9569$ \\
\hline
$10^{-14}$ & $0.25$ & $0.006755$ & $0.025310$ & $3.7468$ \\
\hline
$10^{-14}$ & $0.35$ & $0.007570$ & $0.044008$ & $5.8136$ \\
\hline
$10^{-14}$ & $0.45$ & $0.007745$ & $0.072690$ & $9.3850$ \\
\hline
\end{tabular}
\endgroup
\end{table*}

\begin{figure*}[tp]
	\begin{minipage}{6.5in}
		\centering
		\subfloat[Real parts of the selected reflection zero frequencies.\label{fig:ttm-fixed-radius-spin:a}]{%
			\includegraphics[width=0.485\textwidth]{redraw/figure_13/panel_a.pdf}}\hfill
		\subfloat[Critical wall reflection magnitude.\label{fig:ttm-fixed-radius-spin:b}]{%
			\includegraphics[width=0.485\textwidth]{redraw/figure_13/panel_b.pdf}}
		\par\smallskip
		\includegraphics{redraw/figure_13/legend.pdf}
		\caption{Spin dependence at fixed matching coordinate $z_b$ (circles) and fixed circumferential matching radius $r_b$ (squares).
			Blue solid and magenta dashed curves denote $m=+1$ and $m=-1$, respectively.
			The two configurations coincide at $a=0.6$, indicated by the vertical guide.
			The corotating critical curves exhibit turning points in both cases, but the locations and magnitudes of their sampled minima differ.}
		\label{fig:ttm-fixed-radius-spin}
	\end{minipage}
\end{figure*}
\label{app:branch-checks}

\begin{samepage}
	To replace the abrupt constant-coefficient exterior channel with a smooth lossless connection, let $x_e=x_b+L_{\rm tr}$ and define, for $x_b<x<x_e$,
	\begin{equation}
		\begin{aligned}
			f_{\rm tr}(x)&=[1-S(\xi)]f_{\rm BTZ}(x)+S(\xi)f_b,\\
			S(\xi)&=10\xi^3-15\xi^4+6\xi^5,\\
			\xi&=\frac{x-x_b}{L_{\rm tr}}.
		\end{aligned}
		\label{eq:smooth-completion}
	\end{equation}
	Here $f\in\{N^\varphi,V_m\}$ and $f_b=f_{\rm BTZ}(x_b)$. In the exterior channel at $x\geq x_e$, the coefficients remain fixed at $f_b$. This real $C^2$ interpolation preserves the exterior wave number $k_b$ and introduces no bulk dissipation. The condition $L_{\rm tr}<-x_b$ ensures that the BTZ coefficients remain finite throughout the transition layer. The original abrupt exterior channel is recovered at $L_{\rm tr}=0$.
	\par\end{samepage}

We impose purely incident data $(\Psi,\Psi_x)=(1,-\iu k_b)$ at $x_e$ and integrate inward through the transition layer. Matching to the exact BTZ basis at $x_b$ replaces the coefficient ratio $C_{b,m}$ in the wall inversion by $C_{b,m}^{\rm tr}$. Continuous continuation in $L_{\rm tr}$ tracks the same critical branch, while propagation within the BTZ region remains analytic.

\begin{table}[t]
	\centering
	\small
	\caption{Critical wall reflection magnitudes $10^2|\Kcal|_c$ for the selected branches with the smooth transition layer~\eqref{eq:smooth-completion}.
		The transition length $L_{\rm tr}$ uses the tortoise coordinate units of Eq.~\eqref{eq:ttm-tortoise}, with $L=1$.
		The two sectors have identical critical values in the nonrotating limit.
		All listed positive spin points satisfy $|\Kcal|_{c,+}<|\Kcal|_{c,-}$,
		although the critical values depend on the transition length.}
	\label{tab:smooth-completion}
	\begingroup\GridTableSetup
	\begin{tabular}{|c|c|c|}
		\hline
		$(L_{\rm tr},a)$ & $10^2|\Kcal|_{c,+}$ & $10^2|\Kcal|_{c,-}$ \\
		\hline
		$(0,0)$ & $1.053093$ & $1.053093$ \\
		\hline
		$(0,0.2)$ & $0.574312$ & $1.600511$ \\
		\hline
		$(0,0.4)$ & $0.157557$ & $2.271059$ \\
		\hline
		$(0,0.6)$ & $0.698027$ & $3.069898$ \\
		\hline
		$(0.10,0)$ & $1.141262$ & $1.141262$ \\
		\hline
		$(0.10,0.2)$ & $0.623934$ & $1.718476$ \\
		\hline
		$(0.10,0.4)$ & $0.161313$ & $2.397265$ \\
		\hline
		$(0.10,0.6)$ & $0.709019$ & $3.167992$ \\
		\hline
		$(0.25,0)$ & $1.684690$ & $1.684690$ \\
		\hline
		$(0.25,0.2)$ & $0.963625$ & $2.444124$ \\
		\hline
		$(0.25,0.4)$ & $0.270960$ & $3.210344$ \\
		\hline
		$(0.25,0.6)$ & $0.742402$ & $3.833851$ \\
		\hline
		$(0.40,0)$ & $2.750861$ & $2.750861$ \\
		\hline
		$(0.40,0.2)$ & $1.718104$ & $3.838415$ \\
		\hline
		$(0.40,0.4)$ & $0.706863$ & $4.850478$ \\
		\hline
		$(0.40,0.6)$ & $0.734486$ & $5.307262$ \\
		\hline
	\end{tabular}
	\endgroup
\end{table}

Table~\ref{tab:smooth-completion} shows that the smooth transition layer preserves the nonrotating degeneracy and the sector ordering of the tracked branches at positive spin. At $a=0.6$ and $L_{\rm tr}=0.40$, the critical reflection magnitudes are $0.007344861$ and $0.053072624$ for the two sectors. The $m=+1$ critical value decreases from $L_{\rm tr}=0.25$ to $0.40$, showing that its dependence on the transition length is not monotonic. 

Tightening the transition-layer integration tolerance changes $w_c$ by less
than $7\times10^{-13}$ at $a=0.6$, $L_{\rm tr}=0.40$.
Across Table~\ref{tab:smooth-completion}, the conserved phase flux agrees
with $-k_b$ to within $5\times10^{-13}$ and direct outward propagation gives
$|\Aout/\Ain|<4\times10^{-13}$.

Table~\ref{tab:boundary-position-matching} follows the same critical branches
as $z_w$ and $z_b$ vary from their reference values. The nine configurations
retain the sector ordering, with substantial changes in both critical values
and their ratio.

For the fixed circumferential radius comparison, we set $r_b=2.319151$ and obtain $z_b(a)$ from the inverse radial transformation. Both configurations keep $z_w$ fixed and coincide at $a=0.6$. On a grid with $\Delta a=0.025$, the minimum sampled value of the $m=+1$ critical curve at fixed $z_b$ is $1.504\times10^{-3}$ at $a=0.425$. At fixed $r_b$, the corresponding minimum is $8.83\times10^{-4}$ at $a=0.400$. The corotating turning point persists, but its location and magnitude change. These values are sampled minima only. Neither comparison keeps the full cavity length in tortoise coordinates fixed.

\subsection{Coordinates and exterior matching}
\label{app:control-details}

The scattering calculations and smooth transition layers use the same tortoise coordinate:
\begin{equation}
	\begin{aligned}
		x(r)&=\frac{L^2}{2(r_+^2-r_-^2)}\Bigg[
		r_+\log\frac{r-r_+}{r+r_+}
		\\&\qquad-r_-\log\frac{r-r_-}{r+r_-}\Bigg].
	\end{aligned}
	\label{eq:ttm-tortoise}
\end{equation}
The purely incident condition at the exterior boundary determines the ratio of outgoing to ingoing coefficients in the exact horizon basis:
\begin{equation}
	\begin{aligned}
		C_{b,m}(w)&=-\frac{\mathcal B_b[\Psi_{\rm in}]}
		{\mathcal B_b[\Psi_{\rm out}]},\\
		\mathcal B_b[\psi]&=[\partial_x\psi+\iu k_b(w)\psi]_{x_b}.
	\end{aligned}
	\label{eq:ratio-coefficients}
\end{equation}
The ratio is used where the basis is independent and its denominator is nonzero; $D$ provides the regular characteristic function elsewhere.

The derivative in Eq.~\eqref{eq:logK-leading-law} can equivalently be written
$\dd w/\dd\log\Kcal=[\partial_w\log\Kcal_{{\rm req},m}]^{-1}$.
At $a=0.6$, the dimensional near-horizon phase scale is
$t_{\rm nh}=(L^2/r_+)T_{\rm nh}\simeq21.59$.

\section{Exceptional point structure}
\label{app:reflection-zero-ep}

\begin{table*}[t]
	\centering
	\small
	\caption{Reflection zero EPs obtained by continuation through the
		smooth exterior connections of Eq.~\eqref{eq:smooth-completion}.
		The real parameters $a$ and $\Kcal$ are readjusted at each
		transition length.
		The listed points are connected by continuous continuation at
		intermediate lengths, with the wall remaining real and passive
		throughout.}
	\label{tab:ep-smooth-completion}
	\begingroup\GridTableSetup
	\begin{tabular}{|c|c|c|c|}
		\hline
		$L_{\rm tr}$ & $a_{\rm EP}$ & $10^6\Kcal_{\rm EP}$ & $w_{\rm EP}$\\
		\hline
		$0$ & $0.5919361$ & $-5.973216$ & $0.9074199-0.1359364\iu$\\
		\hline
		$0.10$ & $0.5934565$ & $-6.873984$ & $0.9060937-0.1323521\iu$\\
		\hline
		$0.25$ & $0.6040945$ & $-13.11494$ & $0.8979401-0.1150592\iu$\\
		\hline
	\end{tabular}
	\endgroup
\end{table*}

The EP has $a_{\rm EP}=0.59193611$,
$\Kcal_{\rm EP}=-5.9732165\times10^{-6}$, and
$w_{\rm EP}=0.90741985-0.13593643\iu$.
The cavity-length family holds $r_b/r_+=1.1618050122$ fixed.

For $k_b\ne0$, the unit incident data in
Eq.~\eqref{eq:ep-jost-data} fix
$(\Ain,\Aout)=(1,0)$.
The characteristic function $D=D_0+\Kcal D_1$ therefore vanishes
if and only if this nontrivial radial solution satisfies both
boundary conditions.
In a basis of linearly independent analytic solutions,
$D$ can be expressed as the boundary determinant divided by
the Wronskian.
Its initial-value form avoids poles introduced by ratios of
basis coefficients.
At the EP, $|k_b|=1.27379$, $s_w\ne0$, and $D_1\ne0$,
excluding channel thresholds and degeneracy of the boundary conditions.

Let $\kappa=r_+/L^2$, $s(x,w)=\kappa w+mN^\varphi(x)$,
and $Q(x,w)=s^2-V_m$.
At fixed geometry and wall coefficient, the full nonlinear
eigenvalue problem is
\begin{equation}
	P(w)\psi=\begin{pmatrix}
		(\partial_x^2+Q)\psi\\
		(1+\Kcal)\psi_x(x_w)+\iu(1-\Kcal)s_w\psi(x_w)\\
		\psi_x(x_b)+\iu k_b\psi(x_b)
	\end{pmatrix}=0.
	\label{eq:ep-nonlinear-family}
\end{equation}
For real propagating frequencies, two fields satisfying
the boundary conditions have the Green boundary form
\begin{equation}
	\begin{aligned}
		[\bar u v_x-\bar u_xv]_{x_w}^{x_b}
		&=-2\iu k_b\bar u_bv_b\\
		&\quad+2\iu\operatorname{Re}Z_w\,s_w\bar u_wv_w.
	\end{aligned}
	\label{eq:ep-green-form}
\end{equation}
The nonvanishing boundary form, together with the frequency dependence of the
bulk and boundary equations, makes this a non-self-adjoint nonlinear spectral
problem. For complex $w$, the Green identity also includes the bulk term
proportional to $Q-\bar Q$.

The construction of the nonlinear Jordan chain retains all
frequency dependence.
Writing $s_b=s(x_b,w)$, we have
$k_b'=\kappa s_b/k_b$, $k_b''=-\kappa^2V_b/k_b^3$,
and $\partial_w=\kappa\partial_\omega$.
The corresponding operator derivatives are
\begin{equation}
	\begin{aligned}
		P_w\psi&=\begin{pmatrix}
			2\kappa s\psi\\
			\iu(1-\Kcal)\kappa\psi(x_w)\\
			\iu k_b'\psi(x_b)
		\end{pmatrix},\\[3pt]
		P_{ww}\psi&=\begin{pmatrix}
			2\kappa^2\psi\\0\\\iu k_b''\psi(x_b)
		\end{pmatrix}.
	\end{aligned}
	\label{eq:ep-operator-derivatives}
\end{equation}

The nontrivial outer boundary condition restricts the two-dimensional radial solution space to the span of $j$.
Hence, $D=0$ gives a one-dimensional kernel.
Since $P(w)j(w)=e_{\rm wall}D(w)$,
where $e_{\rm wall}$ selects the inner boundary component,
we set $\psi_0=j(w_{\rm EP})$ and
$\psi_1=\partial_wj(w_{\rm EP})$.
Differentiating at the EP gives a generalized eigenfunction
chain of length two:
\begin{equation}
	\begin{aligned}
		P_0\psi_0&=0,\\
		P_0\psi_1+P_w\psi_0&=0,\\
		\psi_1(x_b)&=0,\qquad
		\psi_{1,x}(x_b)=-\iu k_b'.
	\end{aligned}
	\label{eq:ep-first-chain}
\end{equation}
Here $P_0=P(w_{\rm EP})$, and all operator derivatives are
evaluated at the EP.
The unit incident normalization fixes the freedom to add
a multiple of $\psi_0$ to $\psi_1$.
This nonlinear Jordan chain includes the frequency derivatives
of both boundary conditions
\cite{Guttel:2017nep}.

The next Taylor coefficient $\psi_2=\tfrac12\partial_w^2j$
satisfies
\begin{equation}
	\begin{aligned}
		P_0\psi_2+P_w\psi_1+\tfrac12P_{ww}\psi_0
		&=\tfrac12e_{\rm wall}D_{ww}.
	\end{aligned}
	\label{eq:ep-second-obstruction}
\end{equation}
	This function satisfies the required inhomogeneous radial equation
	and outer boundary condition, but leaves a nonzero wall residual.
	Other solutions satisfying the same radial equation and outer
	boundary condition differ from it only by a multiple of $j$,
	whose wall residual vanishes at the EP because $D=0$.
	If the first generalized eigenfunction is replaced by
	$\psi_1+c\psi_0$, the corresponding second particular solution
	changes by $c\partial_wj$.
	Its additional wall residual is $cD_w=0$.
	Neither freedom can remove the residual of magnitude
	$|D_{ww}|/2=0.2160215$.
	The geometric multiplicity is therefore one, and the Jordan chain
	has length exactly two.

With the original wall normalization, $\Ain\ne0$,
$\Rcal=\Rcal_w=0$, and $\Rcal_{ww}\ne0$,
excluding a trivial vanishing field and cancellation between
a zero and a pole.
Independent inward integration of the radial ODE in $\log z$
agrees with the analytic endpoint field to a relative difference
of $2.77\times10^{-14}$.

The coefficients in Eq.~\eqref{eq:ep-local-unfolding} are
$\beta_a=-0.01554278-0.12147750\iu$ and
$\beta_K=-189.09390-39.63157\iu$.
The derivatives are evaluated at fixed $z_w,z_b$.
The derivative $D_a$ includes changes in the metric,
the endpoints in tortoise coordinates, and the exterior matching data.
Converting from $w$ to $\omega$ multiplies frequency shifts
and their sensitivities by $\kappa$,
while the coefficients in the expansion of the squared frequency
shift are multiplied by $\kappa^2$.
The splitting coefficient agrees with the numerical roots to a relative
error of $6.1\times10^{-7}$ at $\varepsilon=10^{-6}$.

For $L_{\rm tr}>0$, unit incident data are imposed at
$x_e=x_b+L_{\rm tr}$ and propagated inward through the real
$C^2$ transition region, with the same exterior propagation
channel retained.
Table~\ref{tab:ep-smooth-completion} gives the EPs after
readjusting the real parameters $a,\Kcal$.
Independent integration through the transition and BTZ regions
gives $|D|<1.5\times10^{-13}$ and
$|D_w|<2.9\times10^{-12}$ at the continued EPs,
with nonzero $D_{ww}$ and $D_K$.
The EP therefore persists under the tested smooth exterior
connections, although its parameter location shifts.

\begin{figure*}[tp]
	\begin{minipage}{6.5in}
		\centering
		\subfloat[Frequency trajectories as the spin varies.\label{fig:ep-spin-context:a}]{%
			\includegraphics[width=.485\textwidth]{redraw/figure_14/panel_a.pdf}}\hfill
		\subfloat[Imaginary parts of the frequencies.\label{fig:ep-spin-context:b}]{%
			\includegraphics[width=.485\textwidth]{redraw/figure_14/panel_b.pdf}}
		\par\smallskip
		\includegraphics{redraw/figure_14/legend.pdf}
		\caption{Trajectories of the tracked reflection zero branch pair
			as the real spin varies, with $\Kcal=\Kcal_{\rm EP}$ and $m=+1$ fixed.
			Colors and line styles identify the two locally labeled roots,
			and open arrowheads point toward increasing spin, $0\leq a\leq0.8$.
			Open circles and squares mark the actual endpoints at
			$a=0$ and $a=0.8$, respectively.
			The black diamond marks the reflection-zero EP.
			The labels $\sigma=\pm1$ identify the local square-root branches
			on either side of the coalescence.
			Panel~\protect\subref{fig:ep-spin-context:b} shows the imaginary
			parts of the same roots.}
		\label{fig:ep-spin-context}
	\end{minipage}
\end{figure*}
\begin{figure*}[tp]
	\begin{minipage}{6.5in}
		\centering
		\subfloat[Frequency surfaces.\label{fig:ep-control-geometry:a}]{%
			\includegraphics[width=0.35\textwidth]{redraw/figure_15/panel_a.pdf}}\hfill
		\subfloat[Root exchange along a closed loop.\label{fig:ep-control-geometry:b}]{%
			\includegraphics[width=0.315\textwidth]{redraw/figure_15/panel_b.pdf}}\hfill
		\subfloat[Wall sensitivity.\label{fig:ep-control-geometry:c}]{%
			\includegraphics[width=0.315\textwidth]{redraw/figure_15/panel_c.pdf}}
		\par\smallskip
		\includegraphics{redraw/figure_15/legend.pdf}
		\caption{Spectral structure of reflection zeros near the corotating EP.
			Panel~\protect\subref{fig:ep-control-geometry:a} shows the frequency
			surfaces obtained by numerical continuation, with coordinates
			$X=(a-a_{\rm EP})/a_0$ and
			$Y=(\Kcal-\Kcal_{\rm EP})/K_0$,
			where $a_0=9.4232\times10^{-6}$ and
			$K_0=5.9732\times10^{-9}$.
			The green circle is the projection of the parameter loop.
			The chosen branch cut is $Y=0$, $X<0$.
			Panel~\protect\subref{fig:ep-control-geometry:b} follows
			$(a,\Kcal)=(a_{\rm EP}+a_0\cos\theta,\Kcal_{\rm EP}+K_0\sin\theta)$.
			Solid and dashed curves show the real and imaginary parts of
			the frequency shifts, respectively.
			One circuit exchanges the roots, and two circuits restore them.
			Panel~\protect\subref{fig:ep-control-geometry:c} compares the wall
			sensitivity with $C\varepsilon^{-1/2}$, where $C=\tfrac12\sqrt{|\Kcal_{\rm EP}|\,|\beta_K|}\simeq0.016986$.}
		\label{fig:ep-control-geometry}
	\end{minipage}
\end{figure*}

\subsection{Cavity length and zeros of the exterior matching function}
\label{app:ep-length-mechanism}

The family of wall positions in Fig.~\ref{fig:ep-length-control}
keeps $R_b=r_b/r_+$ fixed and connects the exterior channel at
$z_b=(R_b^2-1)/(R_b^2-a^2)$.
The channel coefficients vary with the metric at this fixed
circumferential matching radius.
This family therefore differs from the parameter derivatives at fixed
$z_b$ in Eq.~\eqref{eq:ep-local-unfolding}.

In the region where $D_1\ne0$, the exact characteristic function is
\begin{equation}
	D=D_1(\Kcal-\Kcal_{\rm req}),\qquad
	\Kcal_{\rm req}=-D_0/D_1.
	\label{eq:el-exact-map}
\end{equation}
At a reflection zero, $D_w=-D_1\partial_w\Kcal_{\rm req}$.
Thus, a stationary point of this map corresponds to a double zero
when $\partial_w^2\Kcal_{\rm req}\ne0$.
Near the horizon, the leading form of the map is
\begin{equation}
	\Kcal_{\rm req}\simeq C_{b,+1}(w;a)
	\exp[-\iu T_{\rm nh}(w-aq)].
	\label{eq:el-nh-map}
\end{equation}
Let $w_{\rm ext}(a)$ denote an isolated simple zero of
$C_{b,+1}$ on the selected analytic branch.
At $a=a_{\rm EP}$, this zero is
$w_{\rm ext}=0.90492044-0.11200810\iu$.
Writing $\delta=w-w_{\rm ext}$, we expand
$C_{b,+1}=c\delta(1+b\delta+O(\delta^2))$,
where $c=C_{b,+1}^{\prime}(w_{\rm ext})$
and $b=C_{b,+1}^{\prime\prime}(w_{\rm ext})/(2c)$.

At fixed geometry, the stationary point condition for
Eq.~\eqref{eq:el-nh-map} is
$C_{b,+1}'/C_{b,+1}=1/\delta+b+O(\delta)=\iu T_{\rm nh}$.

The leading frequency and magnitude are given in
Eq.~\eqref{eq:ep-length-leading}; the complex coefficient is
\begin{equation}
 \Kcal_{\rm EP}^{(0)}=-\frac{\iu c}{\mathrm e T_{\rm nh}}
 \exp[-\iu T_{\rm nh}(w_{\rm ext}-aq)].
 \label{eq:ep-length-complex-K}
\end{equation}
Within the linear approximation $C_{b,+1}=c\delta$, the two local roots can
also be written as $\delta=(\iu/T_{\rm nh})W_k(\zeta)$, where
$W_k e^{W_k}=\zeta$ and
$\zeta=-\iu T_{\rm nh}\Kcal\exp[\iu T_{\rm nh}(w_{\rm ext}-aq)]/c$.
The branches $W_0$ and $W_{-1}$ \cite{DLMF:LambertW} meet at $\zeta=-1/\mathrm e$, yielding
$\delta=-\iu/T_{\rm nh}$ and the factor $\mathrm e$ above.
This expression describes the linearized exterior coefficient; curvature
and finite-wall corrections are retained in the numerical roots.

Solving to the next order and substituting into the matching map gives
\begin{align}
	w_{\rm EP}-w_{\rm ext}
	&=-\frac{\iu}{T_{\rm nh}}-\frac{b}{T_{\rm nh}^2}
	+O(T_{\rm nh}^{-3}),\nonumber\\
	\frac{\Kcal_{\rm EP}}{\Kcal_{\rm EP}^{(0)}}
	&=1-\frac{\iu b}{T_{\rm nh}}+O(T_{\rm nh}^{-2}).
	\label{eq:el-curvature-correction}
\end{align}
These expansions require bounded higher Taylor coefficients.
The relevant small parameters include $|b|/T_{\rm nh}$.
The leading curvature correction to $\log|\Kcal_{\rm EP}|$ is
$\operatorname{Im}b/T_{\rm nh}$,
while that to the wall phase is
$-\operatorname{Re}b/T_{\rm nh}$.
A negative real wall coefficient therefore requires
\begin{equation}
	\begin{aligned}
		&\arg c-\frac\pi2-T_{\rm nh}(\operatorname{Re}w_{\rm ext}-aq)\\
		&\quad-\frac{\operatorname{Re}b}{T_{\rm nh}}
		=(2n_{\rm ph}+1)\pi+O(T_{\rm nh}^{-2}).
	\end{aligned}
	\label{eq:el-real-wall-phase}
\end{equation}
The phase is followed continuously along the same analytic branch.

The corresponding amplitude condition is
\begin{equation}
	\begin{aligned}
		\log|\Kcal_{\rm EP}^{(0)}|
		={}&\log|c|-1-\log T_{\rm nh}\\
		&+T_{\rm nh}\operatorname{Im}w_{\rm ext}.
	\end{aligned}
	\label{eq:el-amplitude-balance}
\end{equation}
Along the EP trajectory with a negative real wall coefficient,
the spin must be adjusted with cavity length to satisfy the
phase condition.
This adjustment increases $|c|$ sufficiently to outweigh the
length and damping terms in Eq.~\eqref{eq:el-amplitude-balance},
leading to an increase in the required wall reflection magnitude.
By contrast, at fixed spin with a complex wall coefficient,
the required reflection magnitude decreases from
$1.5910\times10^{-5}$ to $2.3081\times10^{-6}$
over the same range of wall positions.

The near-horizon expansion also permits an additive
$O(z_w)$ correction to $\Kcal_{\rm req}$,
distinct from the Taylor curvature correction to $C_{b,+1}$.
When the wall reflection coefficient is small, this term must
be compared with the exact matching map.
Along the negative real wall EP trajectory, with the spin
readjusted, $|b|/T_{\rm nh}$ changes from $0.112$ to $0.104$.
\subsection{Localized sources and regular response}
\label{app:ep-source-response}

For $\mathcal P_m=(\partial_t-\iu mN^\varphi)^2-\partial_x^2+V_m$,
the retarded Green function satisfies
$\mathcal P_mG_m^R=\delta(t-t')\delta(x-x')$ and vanishes for $t<t'$.
Its frequency-domain equation is
$[-\partial_x^2+V_m-(\omega+mN^\varphi)^2]\widetilde G_m^R=\delta(x-x')$.
If $o$ is the outgoing solution normalized by $o(x_b)=1$ and
$o_x(x_b)=\iu k_b$, then
$\widetilde G_m^R(x,x')=-u_w(x_<)o(x_>)/W[u_w,o]$,
where $W[u_w,o]=u_wo_x-u_{w,x}o$.
Evaluating at the wall, with $u_w(x_w)=1$, gives
$W[u_w,o]=E/(1+\Kcal)$ and recovers
Eq.~\eqref{eq:ep-local-source-green}.
The minus sign supplies the derivative jump $-1$ at the source.
An inverse Fourier--Laplace contour above all frequency singularities
gives the retarded solution.

Set $\Kcal=\Kcal_{\rm EP}+|\Kcal_{\rm EP}|\varepsilon$ and
$\delta\widetilde G_m^R=\widetilde G_m^R(\Kcal)-\widetilde G_m^R(\Kcal_{\rm EP})$.
For identical source and observation positions, the relative complex-field
norm is
\begin{equation}
 \mathcal N_G(\varepsilon)=
 \left[
 \frac{\displaystyle\int_{0.55}^{2.5}
 |\delta\widetilde G_m^R|^2\,\dd w}
 {\displaystyle\int_{0.55}^{2.5}
 |\widetilde G_m^R(\Kcal_{\rm EP})|^2\,\dd w}
 \right]^{1/2}.
 \label{eq:ep-response-norm}
\end{equation}
For $z'=0.05,0.20$ and $x_s=5$, fits over
$10^{-8}\leq|\varepsilon|\leq10^{-3}$ give exponents $0.49998$ and $0.50002$
for the half splitting on opposite sides of the EP and $1$ for
$\mathcal N_G$ at the numerical resolution.
The same norm measures the broadband differences in
Sec.~\ref{sec:reflection-zero-ep}.
Independent radial integration agrees with the exact-basis Green function
to a maximum relative difference of $6.6\times10^{-13}$ at the comparison
points. Sampling details and the additional finite-perturbation comparisons
are retained in the Supplemental Material.

\section{Numerical methods and supplementary checks}
\label{app:numerics}

For the reference parameters in Table~\ref{tab:ttm-parameters},
$q=\pm0.5$, and the endpoints in tortoise coordinates are
$x_w=-11.18431$ and $x_b=-0.74913$.

\subsection{Reflection-zero catalog, continuation, and finite-window coverage}
\label{app:continuation}
\label{app:multimode}

Reflection zeros are roots of $D=0$, followed in $a$ and $\Kcal$ by secant
prediction and Newton correction in 34-digit arithmetic. The eleven pairs
$n=1,\ldots,11$ are ordered by their common frequency at $a=0$ and then
continued without sorting; $n=5$ is the wave-packet pair, and $n=1,2$ in
$m=+1$ coalesce at the EP. The two unpaired displayed branches are denoted
$u^+$ and $u^-$. The reference catalog contains 18 and 16 roots for $m=+1$
and $-1$, respectively; the label correspondence is supplied in the
Supplemental Material.

The argument-principle contours lie on the analytic sheet continued from
$k_b>0$ above the upper channel threshold $w_{\rm th,+}$. Their principal
rectangle is $\max(10^{-6},w_{\rm th,+}+0.03)<\operatorname{Re}w<2.5$,
$|\operatorname{Im}w|<0.3$, excluding the channel cuts, $w=aq$, and
$w=aq\pm\iu(1-a^2)$. Counts agree with the located roots, including
algebraic multiplicity, at both 512 and 1024 intervals per side
(Table~\ref{tab:root-coverage}). The reference rectangles cover 32 of the
34 cataloged roots; two lie in the excluded threshold neighborhoods.
Independent radial integration is compared with the exact solutions in
Supplement~\ref{app:reflection-zero-ep}.

Real-frequency crossings solve
$\operatorname{Im}\Kcal_{{\rm req},m}=0$ with $-1<\Kcal_{{\rm req},m}<0$.
Scans with 501 and 1001 frequencies recover the same 14 and 16 crossings
below $w=2.5$ at $a=0.6$; the lowest in each sector requires
$|\Kcal|_c>0.3$. Such sign-change scans do not exclude tangential zeros.
Figure~\ref{fig:wall-sector-context} shows the wall dependence at $a=0.6$;
the EP in Fig.~\ref{fig:multimode-wall} requires $a=a_{\rm EP}$.

\begin{table}[t]
\centering\small
\caption{Argument-principle counts on the regular scattering sheet.
The main rectangle is defined in the text. At the reference parameters,
additional rectangles above and below the cut supplement the main count.
The nested EP square has half-width $0.01$ and is contained in the main
rectangle. Each count agrees with the located algebraic multiplicity
and is unchanged on doubling the contour sampling.}
\label{tab:root-coverage}
\begingroup\setlength{\tabcolsep}{4pt}
\begin{tabular}{lcc}
\hline
Parameters or region & $m=+1$ & $m=-1$\\
\hline
$(a,\Kcal)=(0,-0.3)$ & 8 & 8\\
$(0.6,-0.3)$ & 13 & 15\\
$(0.8,-0.3)$ & 24 & 29\\
$(0.6,-0.015)$ & 14 & 15\\
$(a_{\rm EP},\Kcal_{\rm EP})$ & 14 & ---\\
EP square & 2 & ---\\
Reference: above cut & 0 & 0\\
Reference: below cut & 4 & 0\\
\hline
\end{tabular}\endgroup
\end{table}

\begin{figure*}[tp]
	\begin{minipage}{6.5in}
		\centering
		\subfloat[$m=+1$.\label{fig:wall-sector-context:a}]{%
			\includegraphics[width=0.485\textwidth]{redraw/figure_16/panel_a.pdf}}\hfill
		\subfloat[$m=-1$.\label{fig:wall-sector-context:b}]{%
			\includegraphics[width=0.485\textwidth]{redraw/figure_16/panel_b.pdf}}
		\par\smallskip
		\includegraphics{redraw/figure_16/legend.pdf}
		\caption{Reflection zero trajectories under variation of the wall
			parameter at the reference spin $a=0.6$, with
			$\Kcal=-|\Kcal|$ and $0.3\geq|\Kcal|\geq10^{-7}$.
			Arrows indicate decreasing wall reflection magnitude,
			and open markers indicate the two sampled wall magnitudes in the legend.
			Small filled circles locate the $|\Kcal|=0.3$ endpoints;
			thin guides connect these endpoints to the branch labels.
			Coral identifies the $n=5$ wave-packet branch.
			Exceptional coalescence at the adjusted spin is shown in
			Fig.~\ref{fig:multimode-wall}.}
		\label{fig:wall-sector-context}
	\end{minipage}
\end{figure*}

\label{sec:port-pseudospectrum}
\label{app:operator-implementation}
\label{subsec:ttm-residual-definition}
\label{subsec:ttm-residual-results}

For the representative zeros at $\Kcal=-0.3$, nonzero
$\partial_w\Rcal$ and independent collocation of the radial equation and
boundary conditions confirm simple roots. Weighted $L^2$ and $H^1$
residuals scale linearly with frequency displacement, with fitted exponents
$0.99561$--$0.99680$ under changes of endpoint weights and numerical floors
of $10^{-11}$--$10^{-12}$. These tests concern the nonlinear spectral
operator \cite{Trefethen:2005,Guttel:2017nep}; an energy-norm resolvent
requires the separate formulation used in
\cite{Jaramillo:2020tuu,Gasperin:2021pseudospectrum,Zhou:2025xdo}.
Details are given in the Supplemental Material.

\subsection{Continuation of the reflection zero EP with wall position}
\label{app:ep-length-validation}

Seventeen converged points along the EP cavity-length trajectory
cover $10\leq-\log_{10}z_w\leq14$ in steps of $0.25$.
At fixed $r_b$, with $\Kcal=-\exp(\log|\Kcal|)$,
we solve $D=D_w=0$ for
$(\operatorname{Re}w,\operatorname{Im}w,a,\log|\Kcal|)$.
All points remain on the same regular analytic branch to the
right of the upper exterior channel threshold.
Along this trajectory,
$|D_{ww}|=0.233$--$0.756$,
$|D_K|=36.6$--$45.8$, and
$|E|=0.105$--$0.247$ remain nonzero.
The real-parameter Jacobian satisfies
$\det[\partial(\operatorname{Re}D,\operatorname{Im}D)/\partial(a,\Kcal)]
\in[-0.879,-0.825]$,
where $D_a=(\partial_aD)_{r_b,z_w}$.

Together with $D_{ww}\ne0$, these results give the nonsingular
real Jacobian required for local continuation of the EP at the
tested points.
Here $D_a$ is defined at fixed circumferential radius of the
exterior matching surface, unlike the derivative at fixed $z_b$
in Eq.~\eqref{eq:ep-local-unfolding}.

With the wall normalization
$\mathcal N_w=|(1+\Kcal)j_x|+|(1-\Kcal)s_wj|$,
independent DOP853 propagation gives
$|D|/\mathcal N_w<5.5\times10^{-16}$ and
$|D_w|/\mathcal N_w<1.2\times10^{-14}$.
The relative difference in $D_{ww}$ is below $1.4\times10^{-11}$.
The propagation retains the bulk and boundary frequency derivatives in
Eq.~\eqref{eq:ep-operator-derivatives}.

The tracked exterior matching zero remains simple,
and both $c$ and the denominator of $C_{b,+1}$ remain nonzero.
The leading prediction for the wall parameter has relative errors
of $0.102$--$0.111$.
Including the correction in Eq.~\eqref{eq:el-curvature-correction}
reduces these errors to $0.0118$--$0.0136$.
By comparison, the difference between the near-horizon matching
map and the exact $\Kcal_{\rm req}$ is below $2.0\times10^{-6}$
at all 17 points.
The curvature term omitted from the leading Taylor expansion
is therefore the main source of error along this trajectory.
Continuation at fixed spin with a complex wall coefficient
separately tests the effect of the real wall phase constraint.

\subsection{Time evolution, fluxes, and convergence}
\label{app:time-numerics}

The wave-packet comparisons use $a=0.6$ and $\Kcal=-0.015$.
The upper exterior channel threshold is $w\simeq0.525519$
for $m=+1$ and $w\simeq0.079294$ for $m=-1$.
The reference packet uses $w_+=1.384580+0.018258\iu$.
The detuned corotating packet uses
$1.458812+0.018258\iu$.
The two counterrotating controls use $w_+$ and
$0.855423+0.018258\iu$, respectively.
All packets retain the same positive growth rate,
with only the oscillation frequency or $m$ sector changed
as described in Sec.~\ref{sec:time-domain}.

\begin{table*}[t]
	\centering
	\small
	\caption{Maximum relative changes across the four wave packets
		for each numerical comparison.
		Percentages are explicitly indicated; all other entries are
		dimensionless ratios.
		The spatial resolution scan uses $N_{\rm grid}=400,560,640$,
		and the time step scan uses $\Delta t=0.06,0.05,0.04$.}
	\label{tab:time-convergence}
	\begingroup\GridTableSetup
	\begin{tabular}{|c|c|c|c|}
		\hline
		Comparison & \GridCell{Reflection\\during incidence}
		& \GridCell{Total\\reflection}
		& \GridCell{Peak energy\\in the region} \\
		\hline
		\GridCell{$N_{\rm grid}=560$\\relative to $640$} & $3.32\%$ & $2.56\%$ & $1.15\times10^{-5}$ \\
		\hline
		\GridCell{$\Delta t=0.05$\\relative to $0.04$} & $1.14\%$ & $1.60\times10^{-7}$ & $6.21\times10^{-6}$ \\
		\hline
	\end{tabular}
	\endgroup
\end{table*}

The time-dependent radial equation and local wall condition are
\begin{align}
	&\left[(\partial_t-\iu mN^\varphi)^2-\partial_x^2+V_m\right]\Psi=0,
	\label{eq:time-radial-equation}\\
	&\left[Z_w(\partial_t-\iu mN^\varphi)-\partial_x\right]
	\Psi\big|_{x_w}=0.
	\label{eq:time-wall-condition}
\end{align}
The coefficients take their exact values in the BTZ region
and remain constant beyond $x_b$.
An outgoing characteristic boundary condition is imposed at
$x_{\max}$.
The wave-packet parameters are
$x_0=10$, $\Delta=4$, $x_c=35$, and $\sigma_x=9$.
Let $G$ be a quintic smooth step function that rises from zero
to unity over $[x_0,x_0+\Delta]$.
The initial field is

\begin{equation}
	\begin{gathered}
		\Psi(0,x)=G(x)
		\begin{cases}
			e^{-\iu k_b(x-x_c)},&x\leq x_c,\\
			\begin{aligned}[t]
				&e^{-(x-x_c)^2/(2\sigma_x^2)}\\
				&\quad\times e^{-\iu\operatorname{Re}k_b(x-x_c)},
			\end{aligned}&x>x_c,
		\end{cases}\\
		\Pi(0,x)=\frac{\omega}{k_b}\partial_x\Psi(0,x).
	\end{gathered}
	\label{eq:finite-vam-pulse}
\end{equation}
Each evolution is normalized by its own initial total energy
$E_{\chi,0}$.

Let $\Pi=\partial_t\Psi$.
The reduced energy and angular momentum densities and fluxes
used in the conservation checks are

\begin{align}
	\rho_E&=\tfrac12\bigl[|\Pi|^2+|\Psi_x|^2\nonumber\\
	&\qquad +(V_m-m^2(N^\varphi)^2)|\Psi|^2\bigr],\nonumber\\
	F_E&=-\operatorname{Re}(\Pi^*\Psi_x),\nonumber\\
	\rho_J&=m\left[-\operatorname{Im}(\Psi^*\Pi)
	+mN^\varphi|\Psi|^2\right],\nonumber\\
	F_J&=m\operatorname{Im}(\Psi^*\Psi_x).
	\label{eq:time-killing-currents}
\end{align}
The energy density and flux associated with the Killing vector
$\chi$ are
$\rho_\chi=\rho_E-\Omega_H\rho_J$ and
$F_\chi=F_E-\Omega_HF_J$.
With $\Omega_w=-N_w^\varphi$, the wall condition gives

\begin{equation}
	F_E-\Omega_wF_J
	=-Z_w\left|(\partial_t-\iu mN_w^\varphi)\Psi_w\right|^2\leq0.
	\label{eq:time-wall-flux}
\end{equation}
The Killing energy flux satisfies the exact relation
$F_\chi=(F_E-\Omega_wF_J)+(\Omega_w-\Omega_H)F_J$.
Net flux conservation and the directional Fourier projections
of Eq.~\eqref{eq:ttm-port-amplitudes} are evaluated separately.

Spatial discretization uses the Chebyshev--Lobatto method
with $N_{\rm grid}+1$ nodes.
Endpoint values are eliminated using the boundary conditions
and their first compatibility conditions.
Time evolution uses a sixth-order Hermite--Pad\'e method.
The reference resolution is $N_{\rm grid}=560$ and $\Delta t=0.05$,
with $x_{\max}=70$ and $t_{\max}=110$.
Diagnostic quantities are recorded every $0.1$ time units.
When $N_{\rm grid}$ changes, barycentric interpolation keeps the
observation point at the fixed physical location $x_s=5$.

The retained propagation band satisfies
$\omega-m\Omega_H>0$ and contains more than $0.999994$
of the field spectral norm.
The stationary spectral prediction discards only components
with amplitudes below $10^{-7}$ of the maximum,
retaining more than $1-1.2\times10^{-12}$ of the incident
$\chi$ flux weight.
Single-frequency tests recover the amplitudes in both
propagation directions to eleven decimal places.
The measured ratios $E_{\chi,{\rm in}}/E_{\chi,0}$ in
Table~\ref{tab:selective-time-controls} are included in
Eq.~\eqref{eq:time-predicted-return-fraction}.

The net flux conservation error in the numerical evolution
is below $2.7\times10^{-5}$ and remains below
$2.84\times10^{-5}$ throughout the convergence scan.
Varying the Tukey taper parameter over
$\alpha_{\rm win}=0,0.04,0.10$ changes the reflected energy during
incidence and the total reflected energy by less than
$5.0\times10^{-6}$ and $3.3\times10^{-5}$ in relative terms,
respectively.
Initially, the energy between the wall and observation point
is below $6.1\times10^{-10}$ of the total initial energy.
Together with Table~\ref{tab:time-convergence},
these checks resolve the ordering of the reflected energies
of the four packets at this resolution.

\phantomsection\label{app:hankel-records}
Hankel--ESPRIT fits \cite{Geissler:2026hankel} to the unfiltered field do
not yield stable QNM frequencies in the available record, $t\leq110$.
For the zero-tuned packet, the fitted dimensionless frequency changes from
$\widehat w=1.38350-0.06912\iu$ on $[70,110]$ to
$1.38231-0.12213\iu$ on $[80,110]$, whereas the nearest locally identified
outgoing pole is $w\simeq1.380682-0.218405\iu$.
We therefore make no QNM identification from these time-domain fits;
the full fitting tests are given in the Supplemental Material.

\FloatBarrier
%